\documentclass[fleqn,usenatbib]{mnras}

\usepackage{mathptmx}

\usepackage[T1]{fontenc}
\usepackage{ae,aecompl}
\usepackage{hyperref}

\usepackage{graphicx}	
\usepackage{amsmath}	
\usepackage{amssymb}	
\usepackage{threeparttable}

\newcommand{\pks}{PKS\,0454$-$22}

\title[\pks]{Discovery and origins of giant optical nebulae surrounding quasar \pks}

\author[Helton et al.]{
Jakob M. Helton,$^{1}$\thanks{E-mail: jhelton@princeton.edu}
Sean D. Johnson,$^{2}$\thanks{Corresponding author; E-mail: seanjoh@umich.edu}
Jenny E. Greene$^{1}$
Hsiao-Wen Chen$^{3}$
\\
$^{1}$Department of Astrophysical Sciences, Princeton University, 4 Ivy Lane, Princeton, NJ 08544, USA\\
$^{2}$Department of Astronomy, University of Michigan, 1085 S. University
323 West Hall, Ann Arbor, MI 48109, USA \\
$^{3}$Department of Astronomy \& Astrophysics, The University of Chicago, 5640 South Ellis Avenue, Chicago, IL 60637, USA
}

\date{Accepted XXX. Received YYY; in original form ZZZ}

\pubyear{2020}

\hypersetup{final}

\begin{document}
\label{firstpage}
\pagerange{\pageref{firstpage}--\pageref{lastpage}}
\maketitle

\begin{abstract}
We report optical integral-field spectroscopy in the field of one of the most luminous quasars in the $z < 1$ Universe, \pks, with the Multi-Unit Spectroscopic Explorer. These data enable the discovery of three large ionized nebulae emitting in [O\,II], H$\beta$, and [O\,III] with projected areas of \textcolor{black}{$1720, \ 1520,$ and $130 \ \mathrm{pkpc}^2$}, which we refer to as N1, N2, and N3, respectively. N1 spatially and kinematically surrounds the quasar host and \textcolor{black}{five} nearby galaxies. The morphology and kinematics of N1 are most consistent with stripped interstellar medium resulting from ongoing interactions. Its ionization properties can be explained by quasar photoionization. N2 spatially and kinematically surrounds two galaxies which are at projected distances of $d \approx 90 \ \mathrm{pkpc}$ and line-of-sight velocities of $\Delta$\textit{v} $\approx +1410\ \mathrm{km\ s^{-1}}$ from the quasar. The morphology and kinematics of N2 are also consistent with stripped interstellar medium. However, its ionization state requires additional ionization sources beyond the quasar, likely from fast shocks as it moves through the hot halo \textcolor{black}{associated with a galaxy over-density around the quasar.} N3 is not coincident with any galaxies with secure redshifts, and may arise from a cool gas structure in the intragroup medium or a dwarf galaxy. These large ionized nebulae demonstrate that interactions can produce cool gas structures on halo scales, while also possibly facilitating quasar fueling. The growing availability of wide-area integral field spectroscopic data will continue to reveal the morphologies, kinematics, and conditions of the gas flows, which may fuel galaxy and black hole growth.
\end{abstract}




\section{Introduction}
\label{sec:1}

Observations have found that supermassive black holes (SMBHs) are present at the centers of most galaxies today \citep[e.g.][]{kormendy_inward_1995, kormendy_coevolution_2013}. Despite the orders-of-magnitude difference in spatial scale, the masses of SMBHs are tightly correlated with various properties of their host galaxy, including the stellar velocity dispersion \cite[e.g.][]{gebhardt_relationship_2000, tremaine_slope_2002} and the spheroid mass \citep[e.g.][]{mclure_black_2002, haring_black_2004}. These tight correlations suggest that SMBHs and galaxy spheroids are formed through a common process. Understanding the physical origins of these relations and their connection to galaxy evolution remains an important, unresolved goal of extragalactic astrophysics. \par

There are a number of related physical processes that may play a role in establishing the observed correlations between SMBHs and their hosts. These include: (1) feedback from actively accreting SMBHs (AGN), (2) common fueling mechanisms, (3) interactions/mergers, and/or (4) group environment. The radiative and mechanical feedback processes associated with AGN can establish the observed correlations between SMBHs and their hosts by removing or heating gas, thus truncating both star formation and black hole growth. Because of this, AGN and their profound energy output have become a crucial part of modern galaxy formation theories \citep[for a review, see][]{somerville_physical_2015}. \par

The observed correlations between SMBHs and their hosts can also be explained by common fueling mechanisms, which may be facilitated by interactions/mergers and galactic environment \citep[e.g.][]{peng_Mergers_2007, jahnke_Scaling_2011}. In particular, fueling nuclear activity requires a redistribution of angular momentum in the interstellar medium (ISM), which allows sufficient gas to infall towards the nucleus \citep[e.g.][]{stockton_qso_1978, yee_quasars_1984, yee_quasars_1987, smith_quasars_1990, bahcall_quasars_correlations_1991, bahcall_quasars_HST_1997, diMatteo_Growth_2003, springel_Mergers_2005, somerville_Coevolution_2008}. While gas-rich galaxy-galaxy interactions/mergers are capable of fueling black hole growth \citep[e.g.][]{barnes_Gasdynamics_1996, hopkins_Fueling_2010}, simulations suggest that their efficacy depends on the host galaxy's environment \citep[e.g.][]{hopkins_unified_2006, hopkins_cosmological_2008, hopkins_cosmological_2008-1} and redshift \citep[e.g][]{McAlpine:2018}. Additionally, observational evidence indicates that black hole growth is most common in overdense galactic regions, such as small groups or clusters, where mergers may be most efficient \citep[e.g][]{hickox_AGN_2009, stott_overdense_2020}. \par

Specifically, both observations \citep{myers_Clustering_2007-1, myers_Clustering_2007-2, ross_Clustering_2009, shen_Clustering_2007, white_Clustering_2012} and theoretical modeling \citep{hopkins_unified_2006, hopkins_cosmological_2008, hopkins_cosmological_2008-1} have shown that AGN are most common in small to intermediate mass groups. There are a few reasons why this characteristic scale may occur. First, a major merger can help redistribute angular momentum and allow ISM to inflow toward the nucleus during a sufficiently short time period \citep[e.g.][]{martini_Lifetimes_2004}. Second, the galaxies involved in this major merger must contain sufficient gas supplies \citep{hernquist_Triggering_1989, barnes_Fueling_1991, barnes_Gasdynamics_1996}. Together, these two effects can cause the characteristic halo mass for gas-rich galaxy-galaxy interactions/mergers, and possibly fueling of the most powerful AGN (quasars; with $L_\mathrm{bol} > 10^{45}\ \mathrm{erg\ s^{-1}}$), to occur in galaxy groups. \par

Understanding the roles of mergers, star formation, and feedback in the coevolution of SMBHs and galaxies requires knowledge of where the gas is located. At $z = 0$, deep 21-cm emission observations can reveal the locations and kinematics of extended but neutral or partially ionized gas \citep[for a review, see][]{duc_Tides_2013}. However, there are almost no luminous quasars to observe at such low redshifts. At $z \gg 0$, there are many luminous quasar to observe, but deep 21-cm observations at these redshifts are not currently possible. Fortunately, luminous quasars and shocks can increase the local ionizing radiation field well above the meta-galactic background \citep[e.g.][]{Haardt:2012, Khaire:2019, Faucher-Giguere:2020}, resulting in photoionization of the denser phases of the circumgalactic medium (CGM). Once photoionized, this gas can emit strongly in rest optical and ultraviolet (UV) emission lines, which are observable over a wide range of redshifts \citep[e.g.][]{Chelouche:2008, Cantalupo_2014, borisova_2016, Epinat:2018, johnson_galaxy_2018, chen_giant_2019}.

\textcolor{black}{Combining this photoionization of large-scale gas supply with the use of wide-field integral field spectroscopy (IFS) provides a powerful way to simultaneously understand the group and gaseous environments of quasars. This combination allows for (1) sensitive observations of extended line-emitting nebulae and (2) joint studies of their morphologies and kinematics in the context of those of nearby galaxies.} Studies like these can determine the properties of quasar host environments, explain the origin of luminous quasars, and improve upon modern galaxy formation theories through a better understanding of the relationship between group environment, large-scale gas supplies, and the evolution of both galaxies and AGN \citep[e.g.][]{husemann_mapping_2010, husemann_muse_2016, johnson_galaxy_2018, chen_giant_2019}. \par

In this paper, we present the discovery of three large ionized nebulae \textcolor{black}{($1720, \ 1520,$ and $130 \ \mathrm{pkpc}^2$ in projected area) emitting strongly in [O\,II], H$\beta$, and [O\,III] using IFS. These nebulae are at projected distances of $d \approx 10-150 \ \mathrm{pkpc}$ from \pks, which is one of the most luminous quasars in the $z < 1$ Universe \citep{Shen:2011}. The morphologies and kinematics of these nebulae alongside} galaxies in the quasar host group show that these nebulae primarily arise from ongoing interactions and cool ($T \approx 10^4 \ \mathrm{K}$) clouds in the intragroup medium (IGrM) rather than large-scale outflows ($\gtrsim 10$ pkpc). These observations provide important insights into the relationship between quasar fueling, gas-rich galaxy-galaxy interactions/mergers, and large-scale gas supplies. \par

This paper proceeds as follows. In Section \ref{sec:2}, we describe the various measurements and observations of \pks. In Section \ref{sec:3}, we present the galactic and nebular environments of this system. In Section \ref{sec:4}, we discuss the origin of the observed nebular photoionization and gas. In Section \ref{sec:5}, we conclude with the implications of our findings. All magnitudes are in the AB system, unless otherwise stated. Throughout the paper, we adopt a standard flat $\Lambda$ cosmology with $\Omega_{\mathrm{m}} = 0.3$, $\Omega_{\Lambda} = 0.7$, and $H_{\mathrm{0}} = 70\ \mathrm{km\ s^{-1}\ Mpc^{-1}}$. \par

\section{Observations and Measurements}
\label{sec:2}

\subsection{\textcolor{black}{Multi-Unit Spectroscopic Explorer Observations}}

Observations by the Multi-Unit Spectroscopic Explorer \citep[MUSE;][]{bacon_muse_2010, bacon_muse_2014} of \pks\ were obtained on January 23, 2018 as part of a European Southern Observatory (ESO) Very Large Telescope (VLT) program (PI: C. P\'eroux, PID: 0100.A-0753) to study the galactic environments of foreground intergalactic medium (IGM) absorption systems at $z \approx 0.27-0.48$, identified in the UV spectra of the quasar \citep{peroux_multi-phase_2019}. The program obtained a total science exposure time of $2700 \mathrm{s}$ under good conditions with median full width at half maximum seeing (FWHM) of 0.6 arcsec. The full MUSE survey and observational strategy will be presented in P\'eroux et al., in prep. \par

MUSE is an optical integral-field spectrograph (IFS) on the VLT UT4 that captures a 1.0 arcmin $\times$ 1.0 arcmin field of view (FOV) with a spatial sampling of 0.2 arcsec $\times$ 0.2 arcsec spaxels in wide-field mode.  Spectrally, MUSE has a contiguous wavelength range of $4750-9350$ \AA\ and resolution of $R \approx 2000-4000$, with higher resolution at the red end. The combination of wide FOV, high spatial sampling, and broad wavelength coverage of MUSE provide an ideal and serendipitous opportunity to study the gaseous and group environment hosting \pks. \par

To study the environment of \pks\, we obtained the publicly-available, reduced MUSE datacube of \pks\ from the ESO Science Archive Facility.  In summary, the ESO MUSE reduction pipeline performs bias subtraction, flat fielding, wavelength calibration, and geometric calibration \citep{weilbacher_design_2012}. All four exposures were reduced independently and then combined to create the final datacube with drizzle resampling. \textcolor{black}{The data are also flux calibrated and sky subtracted using regions free of continuum sources.} The ESO MUSE reduction pipeline uses wavelengths in air, so we converted to vacuum before subsequent analysis. \par

\subsection{Quasar Measurements and Inferred Properties}

While \pks\ is a well studied luminous, radio-loud, jetted quasar \citep[][]{Reid:1999}, the MUSE data offer a new opportunity to examine the spectroscopic properties of the AGN. For \pks, we extracted a 1D MUSE spectrum using \texttt{MPDAF} \textcolor{black}{\citep{piqueras_mpdaf_2017}} with a 1.2 arcsec aperture diameter, chosen to be twice the median seeing of the MUSE datacube. We then measured the redshift from the [O\,II]$\rm \lambda \lambda 3727,3729$ emission line doublet with Gaussian fitting. The best-fit redshift of the quasar is $z_{\rm{QSO}} = 0.5335 \pm 0.0002$. \par

To better understand the intrinsic properties of \pks\ and aid in subsequent photoionization analysis, we also estimated the central black hole mass, bolometric luminosity, and inferred Eddington ratio, following the prescription of \textcolor{black}{\citet{vestergaard_determining_2006}}. We fit the extracted quasar spectrum with a power-law function as the primary continuum, a broadened FeII template spectrum as the secondary continuum \citep{boroson_emission-line_1992, vestergaard_empirical_2001}, and multiple Gaussian line profiles as the emission lines. Prominent broad-line emission is seen in H$\gamma$ and H$\beta$, with three broad Gaussian line profiles being necessary to model these lines. Prominent narrow-line emission is seen in H$\gamma$, H$\beta$, and [O\,III]$\lambda\lambda 4960,5008$, with two narrow Gaussian line profiles being necessary to model these lines. We inferred a \textcolor{black}{continuum monochromatic luminosity ($\lambda L_{\lambda}$)} at $5100$ {\AA} (rest frame) of \textcolor{black}{$\lambda L_{\rm{5100}} \approx 10^{46}\ \rm{erg\ s^{-1}}$} and a bolometric luminosity of $L_{\rm{bol}} \approx 10^{47}\ \rm{erg\ s^{-1}}$, with bolometric corrections from \cite{richards_spectral_2006}. \textcolor{black}{This luminosity means \pks\ is among the most luminous 0.1\% of quasars at $z < 1$, when compared to the Sloan Digital Sky Survey (SDSS) Data Release 7 (DR7) quasar catalog \citep[][]{Shen:2011}.} Using an empirical relationship between the line width of the H$\beta$ broad component and optical continuum luminosity, calibrated with mass measurements of local AGN based on emission line reverberation mapping \textcolor{black}{\citep{vestergaard_determining_2006}}, we estimate a fiducial virial mass from H$\beta$ of $M_{\mathrm{BH}} \approx 4 \times 10^{9}\ M_{\odot}$. Together, the bolometric luminosity and central black hole mass give an inferred Eddington ratio of $\lambda \approx 0.16$. The quasar spectrum used to make these measurements is shown in the top panel of Figure \ref{fig:spectrum_galaxies_1}. \par

\subsection{Quasar Light Subtraction}

\pks\ is a bright quasar with $V \approx 16$ mag resulting in significant contamination of sources within $\lesssim$ 6 arcsec from the quasar due to broad wings on the MUSE point-spread function (PSF), even under good seeing conditions. To remove this contaminating flux, we developed a quasar light subtraction technique that is free of assumptions about the PSF shape, \textcolor{black}{and instead uses spectral information from the IFS and the fact that galaxy and quasar spectral energy distributions are significantly different \citep[see also][]{rupke_quasar-mode_2017}. Because the PSF is more (less) extended at bluer (redder) wavelengths, spectra extracted close to (far from) the quasar centroid appear flatter (steeper) as shown in Figure \ref{fig:appendix1}.} We used non-negative matrix factorization (NMF) to account for this \textcolor{black}{wavelength dependence} \citep{blanton_k_2007, ren_non-negative_2018}, which is described in detail in the appendix and summarized here. \textcolor{black}{We performed two-component NMF on quasar-dominated spaxels from a 1.0 arcsec $\times$ 1.0 arcsec aperture centered around the quasar. The first non-negative spectral component has a steep spectral slope and approximates quasar contamination far from the quasar, while the second non-negative spectral component has a shallow spectral slope and approximates quasar contamination close to the quasar.} Varying the relative contribution of these two components enables effective subtraction of the quasar contamination by accounting for the wavelength dependence of the PSF. We modeled each \textcolor{black}{individual} spaxel at $\lesssim$ 6 arcsec from the quasar as a linear combination of the two quasar spectral components and the first two SDSS galaxy eigenspectra from the Baryon Oscillation Spectroscopic Survey \citep[BOSS;][]{bolton_spectral_2012}. In order to be most sensitive to galaxies at the quasar redshift, we shifted the BOSS eigenspectra to $z_{\rm{QSO}} =  0.5335$ with the strongest nebular emission lines masked ([O\,II], H$\beta$, and [O\,III]). \par

We then subtracted the quasar component of the best-fit model from each spaxel. \textcolor{black}{This effectively removes the quasar light contribution at $\gtrsim$ 1 arcsec from the quasar \citep{johnson_galaxy_2018}.} At $\lesssim$ 1 arcsec from the quasar centroid, the residuals are significant as seen in Figure \ref{fig:appendix2}. We masked this region in subsequent analysis. \par

\subsection{\textcolor{black}{Galaxy Detection, Measurements, and Inferred Properties}}
\label{sec:2.4}

Galaxy surveys in the field of \pks\ have been conducted previously in order to study the environments of foreground absorbers \citep{Chen1998, Chen2001}. These surveys identified eight galaxies, two of of which are at the quasar redshift, but the new MUSE data enable significantly deeper and more complete surveys within its 1.0 arcmin $\times$ 1.0 arcmin FOV. \par

We identified \textcolor{black}{detected objects} in the field of \pks\ with \texttt{Source Extractor} \citep{bertin_sextractor:_1996} after performing quasar light subtraction.  To do this, we used both a broad-band image created from the MUSE datacube with spectral coverage of $6250-7250$ {\AA} \textcolor{black}{(which does not include strong emission lines at $z_{\rm{QSO}} = 0.5335$)} and an archival image from the Wide Field and Planetary Camera 2 (WFPC2) on the \textit{Hubble Space Telescope} (HST) with the F702W filter and total integration time of $4800 \mathrm{s}$ (PI: K. Lanzetta, PID: 7475). We noticed a number of sources visible in the broad-band and HST images that were not identified by \texttt{Source Extractor} due to a spatially dependent background that is not effectively captured by the \texttt{Source Extractor} background estimation routine. The spatial variation in the background is primarily the result of bright point sources in the field as well as detector artifacts. We included these manually identified sources in our continuum source catalog, with centroids measured from local 2D Gaussian fits. The resulting catalog is complete at $m_{\mathrm{F702W}} \approx 25 - 26$ based on background sky estimation from the HST image and the WFPC2 exposure time calculator. \par

For each detected continuum source from \texttt{Source Extractor}, we extracted the corresponding 1D MUSE spectrum using \texttt{MPDAF} \textcolor{black}{\citep{piqueras_mpdaf_2017}} with aperture diameters determined by \texttt{Source Extractor}. These aperture diameters represent the minimum spatial root mean square (rms) dispersion of the object profile along any direction. For each manually detected continuum source, we extracted the corresponding 1D MUSE spectrum using \texttt{MPDAF}  with 1.2 arcsec aperture diameters. These aperture diameters are chosen to be twice the median seeing of the MUSE datacube. In some cases, a smaller aperture is chosen to mitigate crowding effects. \par

We then measured the redshift of each source by determining the best-fit redshift using SDSS galaxy eigenspectra from BOSS. At each redshift on a grid from $z = 0.0$ to $z = 1.0$ with steps of $\Delta z = 0.0001$, we fit the observed spectrum with a linear combination of the SDSS eigenspectra. \textcolor{black}{To do this fit, we explicitly solve matrix equations to find the best-fit linear combination of PCA models using \texttt{NumPy} linear algebra functions.} We adopted the global minimum-$\chi^{2}$ solution as our initial redshift measurement, visually inspected the resulting $\chi^{2}$ grid, and compared the observed spectrum to the best fit in order to judge redshift measurement quality. To ensure robust redshift measurements, we required at least two distinct observed spectral features to be fit properly. In some cases, prominent emission features occur at velocities that are distinct from the stellar absorption features due to relative velocities between galaxies and spatially coincident nebular emission. For examples of this, see G2 in Figure \ref{fig:spectrum_galaxies_1} and G13/G14 in Figure \ref{fig:spectrum_galaxies_2}, which exhibit stellar absorption velocities offset by $\Delta$\textit{v} $= -820\, , \ -1530\, , \ -1390 \ \mathrm{km\ s^{-1}}$ from the nebular emission velocities, respectively. In these cases, we masked the strong emission lines so that the redshift measurements are driven by stellar absorption features. This ensures that the measured redshift represents that of the galaxy rather than extended nebulae. \textcolor{black}{The models shown in these figures are solely used for redshift measurements and are included to emphasize the offset between galactic and nebular velocities in the cases where strong emission lines needed to be masked (G2/G13/G14).} The resulting galaxy redshift uncertainties correspond to velocity uncertainties of $\approx$ 20 $\rm km\ s^{-1}$, though in some cases the errors are larger due to low S/N in broad stellar absorption features. This redshift survey is approximately $100$\% complete at $m_{\mathrm{F702W}} \approx 23$ and approximately $90$\% complete at $m_{\mathrm{F702W}} \approx 24$ (which corresponds to $M_{g} \approx -18.5$ at $z_{\rm{QSO}}$ = 0.5335) \textcolor{black}{based on the redshift measurement success rate and the completeness of continuum source catalog.} The high completeness of the galaxy detection and redshift measurements enable a detailed investigation of possible environmental dependence of extended gas around galaxies. \par

\textcolor{black}{To estimate the stellar masses of the galaxies in the quasar host environment, we performed stellar population synthesis (SPS) with \texttt{Bagpipes} \citep{Carnall:2018}. In summary, \texttt{Bagpipes} uses the \texttt{MultiNest} \citep{Feroz:2013, Buchner:2014} Bayesian inference package to find the best-fit star formation history and associated uncertainty using the stellar evolution models from \citet{bruzual_stellar_2003} assuming an initial mass function from \citet{Kroupa:2003}. To perform the SPS, we assume an exponentially declining star formation history with $0.01 < \tau/\mathrm{Gyr} < 5$, stellar metallicity with $0.2 < Z/Z_{\odot} < 2.5$, and a \citet{Calzetti:2000} dust law with $0 < A_{V}/\mathrm{mag} < 2$. For each galaxy, we performed the SPS fit to the HST photometry and MUSE spectrum simultaneously with the redshift allowed to vary by $\pm$ 600 $\rm km\ s^{-1}$ from the best-fit redshift measured from the PCA fit. In addition, the fitting includes a low-order multiplicative polynomial, which is intended to absorb any flux calibration errors in the MUSE spectrophotometry. The resulting estimated stellar masses marginalized over all other parameters are reported in Table \ref{tab:summary_galaxies} and have typical uncertainties of $\approx 0.2$ dex. We caution that the stellar mass inferred for G1 is subject to unusually large and difficult to quantify systematic uncertainty due to quasar light contaminating both the MUSE spectrum and HST photometry.} \par

\textcolor{black}{In addition to stellar masses, the \texttt{Bagpipes} stellar population synthesis models provide a consistency check on our PCA-based redshift estimates. We find good agreement between the two redshift measurement techniques with a median difference of $\approx$ 10 $\rm km\ s^{-1}$ and a $1\sigma$ scatter of $\approx$ 20 $\rm km\ s^{-1}$ for galaxies with sufficient S/N in stellar absorption features. For galaxies G1, G3, G10, and G19, the stellar continuum data quality is insufficient to measure a redshift independent of emission lines, and for galaxies G7 and G18, low S/N result in a somewhat larger redshift uncertainty of $\approx$ 100 $\rm km\ s^{-1}$.} \par

\section{The Environment of PKS0454$-$22}
\label{sec:3}

\subsection{\textcolor{black}{Galaxy Measurements and Inferred Properties}}

\begin{figure*}
	\includegraphics[width=\textwidth]{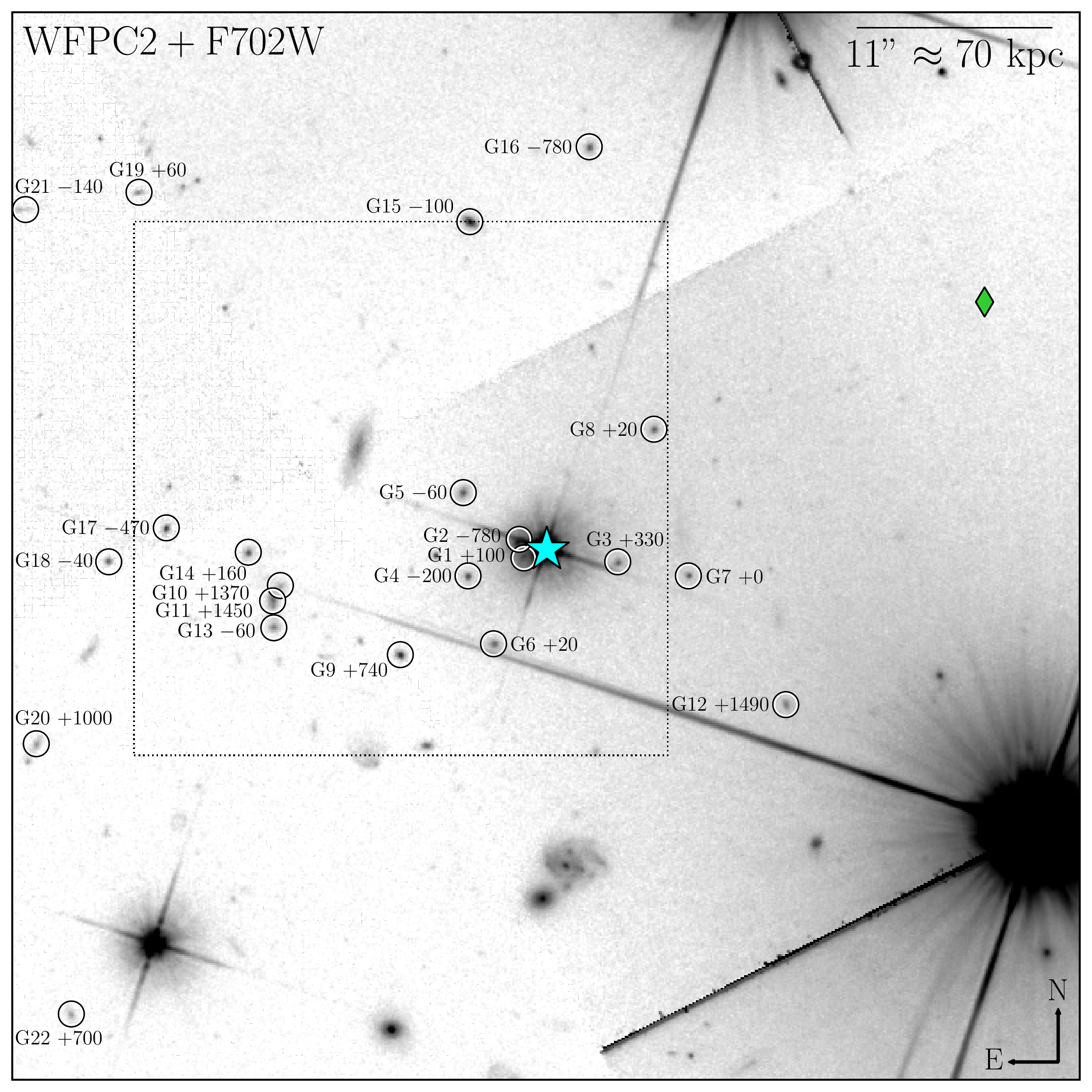}
    \caption{HST WFPC2+F702W image of the field of \pks. The full image shows the 1.0 arcmin $\times$ 1.0 arcmin MUSE FOV and the dotted box marks the 30 arcsec $\times$ 30 arcsec region displayed in Figure \ref{fig:velocity_panels}. Galaxies in the quasar host environment are labeled by their ID and LOS velocity from the quasar in $\mathrm{km\ s^{-1}}$ ($z_{\rm{QSO}} = 0.5335$). The quasar is marked by a cyan star while one of the quasar radio lobes is marked by a green diamond. The other radio lobe falls outside of the MUSE FOV, so it is not shown here.}
    \label{fig:HST_image}
\end{figure*}

\begin{figure*}
	\includegraphics[width=0.90\textwidth]{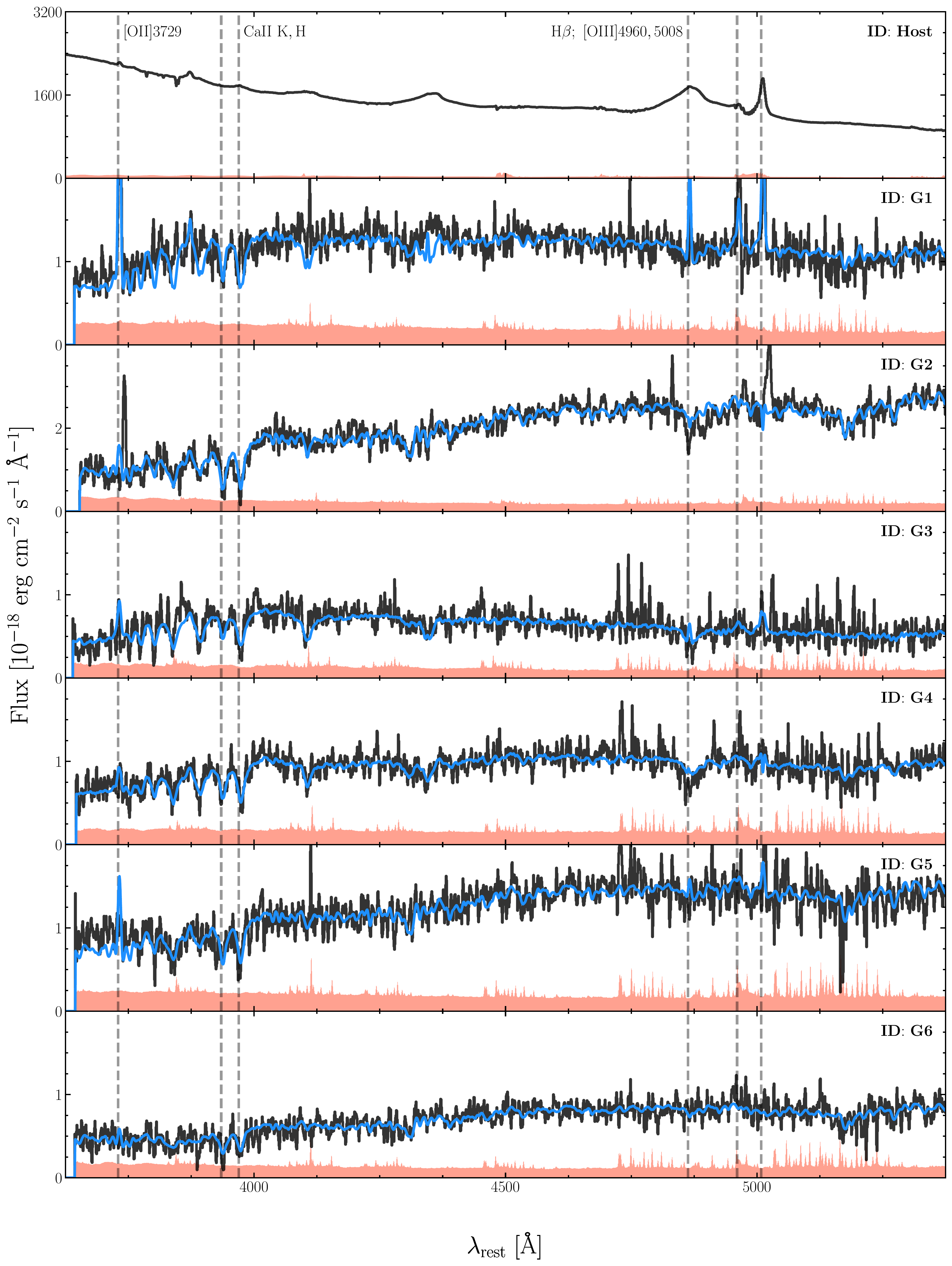}
    \caption{MUSE quasar and galaxy spectra and best-fitting spectral models for the 7 closest galaxies (including the quasar) that fall within the 30 arcsec $\times$ 30 arcsec region displayed in Figure \ref{fig:velocity_panels}. The MUSE spectrum is shown by a solid line in black with errors shown by a filled area in red, while the best-fit spectral model is shown by a solid line in blue. This best-fit spectral model is solely used for redshift measurements. In some cases, prominent emission features occur at velocities that are distinct from stellar absorption features. In these cases, we masked the strong emission lines so that the redshift measurements are driven by stellar absorption features. Galaxy labels are in the upper-right corner of each panel. }
    \label{fig:spectrum_galaxies_1}
\end{figure*}

\begin{figure*}
	\includegraphics[width=0.90\textwidth]{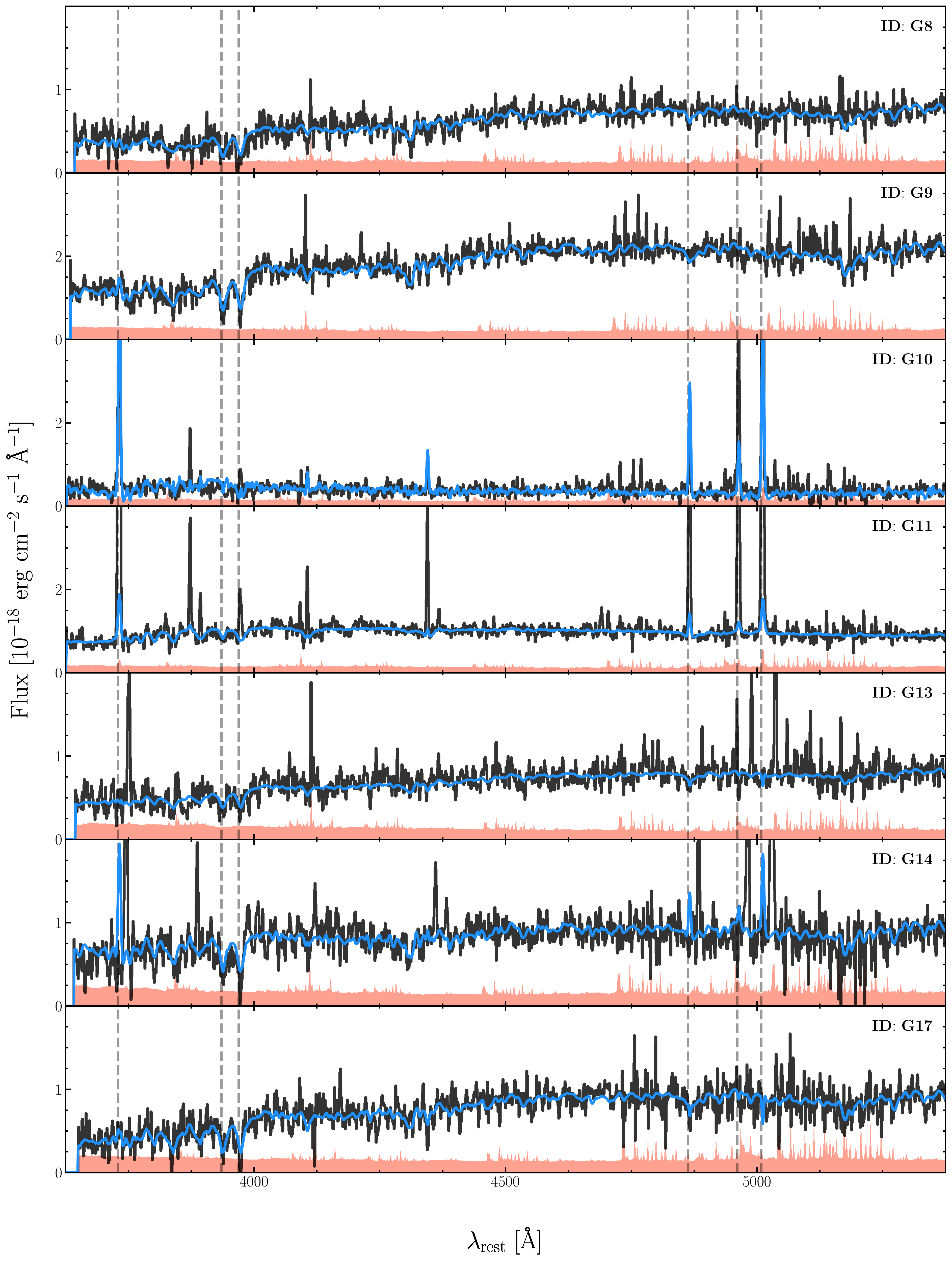}
    \caption{MUSE galaxy spectra and best-fitting spectral models for the 7 furthest galaxies that fall within the 30 arcsec $\times$ 30 arcsec region displayed in Figure \ref{fig:velocity_panels}. The MUSE spectrum is shown by a solid line in black with errors shown by a filled area in red, while the best-fit spectral model is shown by a solid line in blue. This best-fit spectral model is solely used for redshift measurements. In some cases, prominent emission features occur at velocities that are distinct from stellar absorption features. In these cases, we masked the strong emission lines so that the redshift measurements are driven by stellar absorption features. Galaxy labels are in the upper-right corner of each panel. }
    \label{fig:spectrum_galaxies_2}
\end{figure*}

\begin{figure*}
	\includegraphics[width=\textwidth]{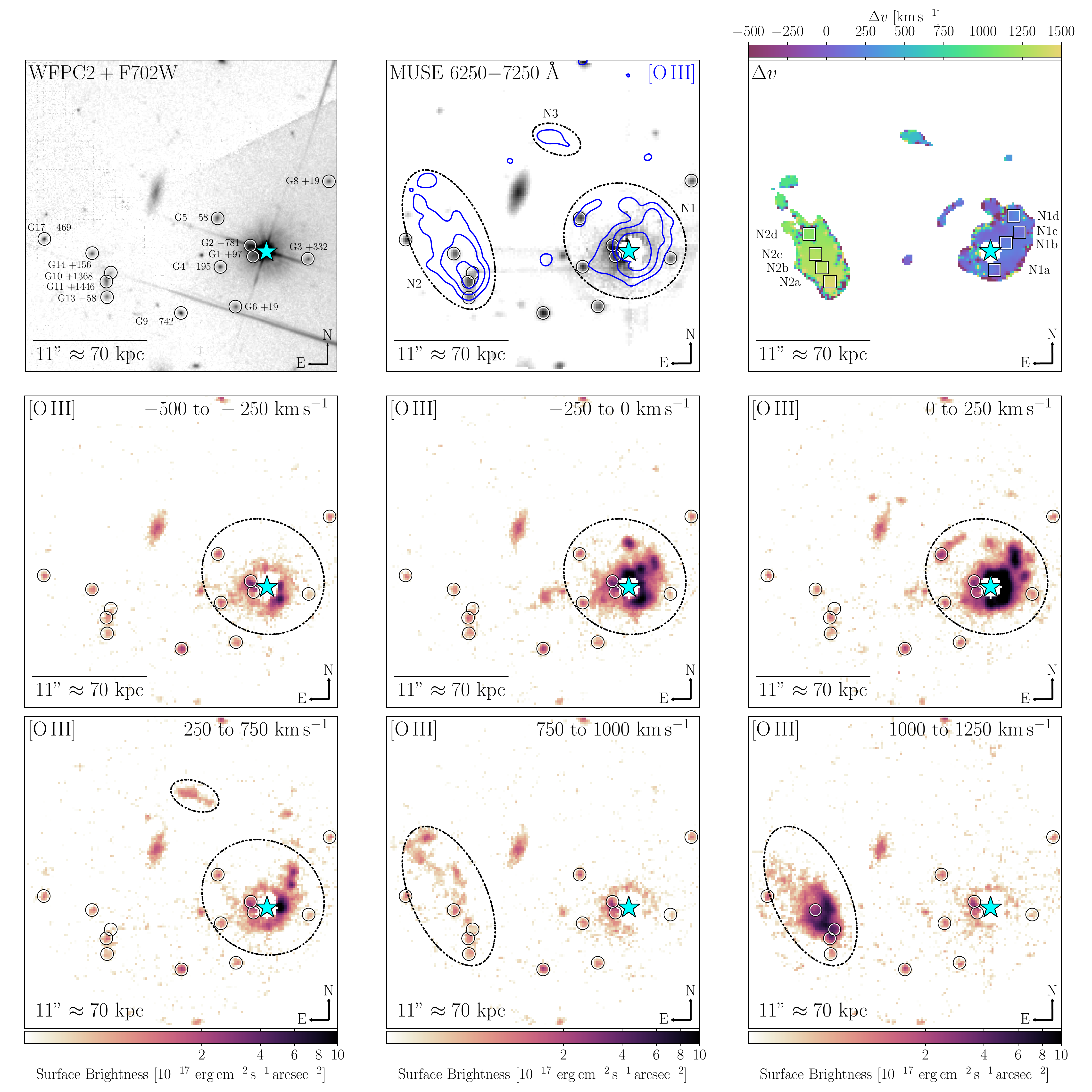}
    \caption{\textit{Top-left panel}: HST WFPC2+F702W image of the field of \pks. \textcolor{black}{Galaxies in the quasar host environment are marked by circles and labeled by their IDs and LOS velocities relative to the quasar.} \textit{Top-middle panel}: median flux image over $6250-7250$ {\AA} in the MUSE datacube with [O\,III] surface brightness contours of 0.2, 0.8, and 3.2 $\times\ 10^{-17}\ \mathrm{erg\ cm^{-2}\ s^{-1}\ arcsec^{-2}}$ in blue. Nebulae are marked by dashed ellipses and labeled by their IDs, the positions of galaxies in the host group are marked by circles, and the quasar is marked by a cyan star.  \textit{Top-right panel}: map of the nebular LOS velocities relative to the quasar, showing spaxels with significant detection of [O\,III] (S/N > 3). \textcolor{black}{Galaxies in the host group are marked by circles, extracted nebular regions are marked by squares and labeled by their IDs, and the quasar is marked by a cyan star. \textit{Middle row and bottom row}: narrow-band [O\,III] images extracted from the MUSE datacube over the given LOS velocity interval. Galaxies in the host group are marked by circles, nebulae are marked by dashed ellipses, and the quasar is marked by a cyan star.}}
    \label{fig:velocity_panels}
\end{figure*}

\textcolor{black}{Candidate members of the quasar host group were selected by implementing a line-of-sight (LOS) velocity cut (|$\Delta$\textit{v}| $< 2000\ \rm km\ s^{-1}$ from the quasar) on galaxies with secure redshifts from Section \ref{sec:2.4}.} We chose this velocity range to be approximately twice the velocity dispersion of the most massive galaxy clusters. \textcolor{black}{There are 23 galaxies in the MUSE field within the given velocity range, including the quasar host.} Figure \ref{fig:HST_image} shows a 1.0 arcmin $\times$ 1.0 arcmin cutout of the HST WFPC2+F702W image of the field, with these group members labeled by their ID and LOS velocity from the quasar. This cutout matches the 1.0 arcmin $\times$ 1.0 arcmin MUSE FOV. We also mark one of the quasar radio lobes with a green diamond, while the other radio lobe falls outside of the image \citep{Reid:1999}. Observed and inferred properties of these galaxies are shown in Table \ref{tab:summary_galaxies}. Additionally, Figures \ref{fig:spectrum_galaxies_1} and \ref{fig:spectrum_galaxies_2} shows the MUSE galaxy spectra and the best-fitting spectral models used for redshift measurements for group members that fall within the 30 arcsec $\times$ 30 arcsec region displayed in Figure \ref{fig:velocity_panels}. \par

\begin{table*}
	\centering
	\caption{Summary of Galaxies in the Field of \pks\ at \textit{$z \approx z_{\rm{QSO}}$}.}
	\label{tab:summary_galaxies}
	\begin{threeparttable}
	\begin{tabular}{cccccccrrrr} 
		\hline
		ID & R.A.\tnote{a} & Decl.\tnote{b} & \textit{z}\tnote{c} & \textit{$m_{\mathrm{F702W}}$}\tnote{d} & \textit{$M_{g}$}\tnote{e} & \textit{u-g}\tnote{f} & \multicolumn{1}{c}{$\log (M_*/M_\odot)$\tnote{g}} & \multicolumn{1}{c}{$\Delta \theta$\tnote{h}} & \multicolumn{1}{c}{\textit{d}\tnote{i}} & \multicolumn{1}{c}{$\Delta$\textit{v}\tnote{j}} \\
		& (J2000) & (J2000) & & (AB) & (AB) & (AB) & & (arcsec) & (pkpc) & ($\mathrm{km\ s^{-1}}$) \\
		\hline
		Host & 04:56:08.90 & \textcolor{black}{$-$21:59:09.1} & 0.5335 &  ... &     ... & ... &  ... &  0.0 &   0.0 & 0\\
		 G1  & 04:56:08.99 & \textcolor{black}{$-$21:59:09.8} & 0.5340 & 23.2 & $-$19.3 & 0.3 & \textcolor{black}{10.6} &  1.5 &   9.9 & +100\\
		 G2  & 04:56:09.01 & \textcolor{black}{$-$21:59:08.7} & 0.5295 & 22.5 & $-$20.0 & 0.8 & \textcolor{black}{10.8} &  1.7 &  11.0 & -780\\
		 G3  & 04:56:08.61 & \textcolor{black}{$-$21:59:10.0} & 0.5352 & 23.8 & $-$18.6 & 0.2 & \textcolor{black}{ 8.9} &  4.4 &  28.1 & +330\\
		 G4  & 04:56:09.22 & \textcolor{black}{$-$21:59:10.8} & 0.5325 & 23.4 & $-$19.1 & 0.4 & \textcolor{black}{10.1} &  5.1 &  32.4 & -200\\
		 G5  & 04:56:09.24 & \textcolor{black}{$-$21:59:06.1} & 0.5332 & 23.1 & $-$19.5 & 0.6 & \textcolor{black}{10.8} &  5.9 &  37.7 & -60\\
		 G6  & 04:56:09.12 & \textcolor{black}{$-$21:59:14.6} & 0.5336 & 23.7 & $-$18.9 & 0.7 & \textcolor{black}{10.4} &  6.4 &  40.8 & +20\\
		 G7  & 04:56:08.33 & \textcolor{black}{$-$21:59:10.8} & 0.5335 & 23.9 & $-$18.5 & 0.3 & \textcolor{black}{ 9.4} &  8.7 &  56.0 & +0\\
		 G8  & 04:56:08.47 & \textcolor{black}{$-$21:59:02.5} & 0.5336 & 23.8 & $-$18.8 & 0.8 & \textcolor{black}{10.6} &  9.2 &  59.2 & +20\\
		 G9  & 04:56:09.49 & \textcolor{black}{$-$21:59:15.2} & 0.5373 & 22.6 & $-$19.9 & 0.7 & \textcolor{black}{10.6} & 10.8 &  69.1 & +740\\
		G10  & 04:56:09.98 & \textcolor{black}{$-$21:59:11.3} & 0.5405 & 24.2 & $-$18.3 & 0.3 & \textcolor{black}{ 9.3} & 16.3 & 104.5 & +1370\\
		G11  & 04:56:10.01 & \textcolor{black}{$-$21:59:12.2} & 0.5409 & 23.2 & $-$19.4 & 0.4 & \textcolor{black}{ 9.7} & 16.9 & 108.3 & +1450\\
		G12  & 04:56:07.93 & \textcolor{black}{$-$21:59:18.0} & 0.5411 & 23.7 & $-$18.8 & 0.2 & \textcolor{black}{ 9.5} & 17.0 & 109.0 & +1490\\
		G13  & 04:56:10.00 & \textcolor{black}{$-$21:59:13.7} & 0.5332 & 23.7 & $-$18.9 & 0.6 & \textcolor{black}{10.0} & 17.2 & 110.0 & -60\\
		G14  & 04:56:10.11 & \textcolor{black}{$-$21:59:09.5} & 0.5343 & 23.4 & $-$19.1 & 0.5 & \textcolor{black}{ 9.8} & 18.1 & 116.0 & +160\\
		G15  & 04:56:09.21 & \textcolor{black}{$-$21:58:50.9} & 0.5330 & 22.1 & $-$20.4 & 0.8 & \textcolor{black}{10.5} & 18.8 & 120.6 & -100\\
		G16  & 04:56:08.73 & \textcolor{black}{$-$21:58:46.7} & 0.5295 & 23.4 & $-$19.2 & 0.7 & \textcolor{black}{10.2} & 22.6 & 144.7 & -780\\
		G17  & 04:56:10.44 & \textcolor{black}{$-$21:59:08.1} & 0.5311 & 23.6 & $-$19.0 & 0.7 & \textcolor{black}{10.4} & 23.1 & 147.9 & -470\\
		G18  & 04:56:10.67 & \textcolor{black}{$-$21:59:10.0} & 0.5333 & 23.4 & $-$19.1 & 0.6 & \textcolor{black}{10.2} & 26.6 & 170.3 & -40\\
		G19  & 04:56:10.55 & \textcolor{black}{$-$21:58:49.2} & 0.5338 & 24.0 & $-$18.4 & 0.2 & \textcolor{black}{ 9.2} & 31.7 & 203.3 & +60\\
		G20  & 04:56:10.96 & \textcolor{black}{$-$21:59:20.2} & 0.5386 & 24.1 & $-$18.4 & 0.5 & \textcolor{black}{ 9.6} & 32.9 & 210.7 & +1000\\
		G21  & 04:56:11.01 & \textcolor{black}{$-$21:58:50.2} & 0.5328 & 23.7 & $-$18.7 & 0.4 & \textcolor{black}{ 9.6} & 36.8 & 236.0 & -140\\
		G22  & 04:56:10.82 & \textcolor{black}{$-$21:59:35.4} & 0.5371 & 24.2 & $-$18.2 & 0.3 & \textcolor{black}{ 9.8} & 39.0 & 250.0 & +700\\
		\hline \\
	\end{tabular}
	\begin{tablenotes}
	    \footnotesize
        \item \textbf{Notes.}
        \item[a] Right ascension.
        \item[b] Declination.
        \item[c] \textcolor{black}{Best-fit redshift, from principal component analysis of SDSS galaxy eigenspectra from BOSS.}
        \item[d] Apparent HST WFPC2+F702W magnitude, \textcolor{black}{in the AB system.}
        \item[e] Absolute \textit{g-}band magnitude, \textcolor{black}{in the AB system.}
        \item[f] Rest-frame \textit{u-g} color, measured in matched isophotal apertures determined by \texttt{Source Extractor}.
        \item[g] \textcolor{black}{Stellar mass, from stellar population fit to the MUSE spectrum and HST photometry.}
        \item[h] Angular distance from the quasar, with units of arcsec.
        \item[i] Projected physical distance from the quasar, with units of pkpc. 
        \item[j] LOS velocity from the quasar, with units of $\mathrm{km\ s^{-1}}$.  
    \end{tablenotes}
	\end{threeparttable}
\end{table*}

\begin{figure}
    \centering
	\includegraphics[width=1.0\linewidth]{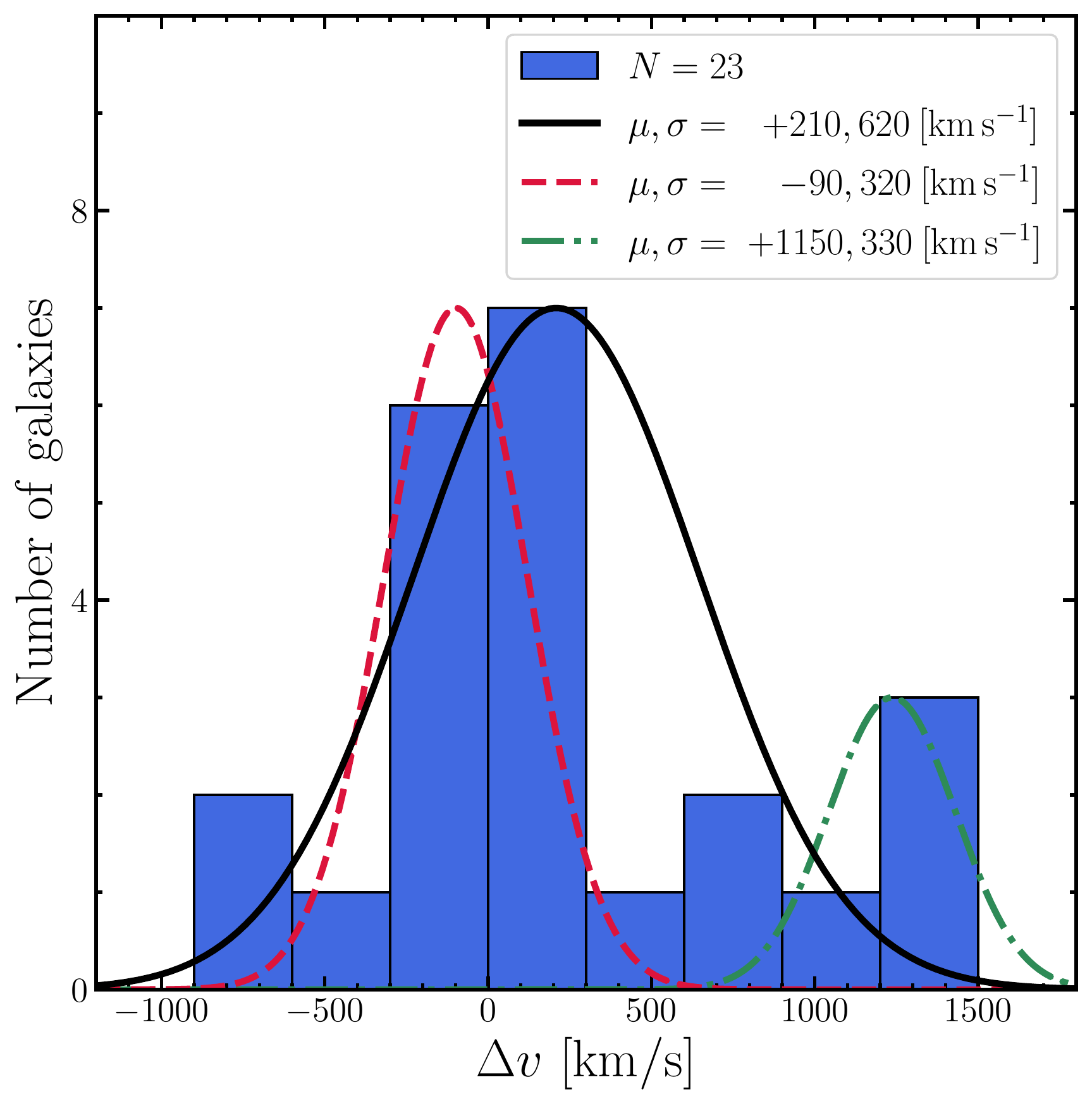}
    \caption{Histogram displaying the LOS velocity from the quasar ($\Delta$\textit{v} = \textit{v} - \textit{v}$_{\mathrm{QSO}}$) of galaxies in the galactic environment of \pks. The best-fit single-Gaussian to the full distribution is shown by a solid line in black, while the best-fit \textcolor{black}{two-component Gaussian Mixture Model} is shown by a dashed line in red and a dash-dotted line in green. These two sets of Gaussian fits were performed to place upper (single-component Gaussian) and lower (two-component Gaussian) limits on the velocity dispersion and mass of the quasar host group.}
    \label{fig:histogram}
\end{figure}

The quasar host environment includes 10 (23) galaxies of $m_{\mathrm{F702W}} \leq 23.5$ ($24.5$) which suggests a relatively massive group. The distribution of galaxy LOS velocities relative to the quasar ($\Delta$\textit{v} = \textit{v} - \textit{v}$_{\mathrm{QSO}}$) is shown in Figure \ref{fig:histogram}. This distribution is non-Gaussian, with a peak near the quasar systemic velocity ($\Delta$\textit{v} $\approx +0\ \mathrm{km\ s^{-1}}$) and a significant tail at \textcolor{black}{$\Delta$\textit{v} $\gtrsim +700\ \mathrm{km\ s^{-1}}$}. Galaxies with velocities in the peak near the quasar systemic velocity fall on all sides of the quasar while those in the tail at \textcolor{black}{$\Delta$\textit{v} $\gtrsim +700\ \mathrm{km\ s^{-1}}$} are more typically found $\gtrsim 50 \ \mathrm{pkpc}$ to the east of the quasar. This may be explained if the quasar host group is in the process of interacting/merging with a second group of galaxies. \par

To characterize the velocity dispersion and mass(es) of the group(s), we performed two sets of Gaussian fits to the distribution of galaxy LOS velocities. We first fit a single Gaussian to the full distribution, finding a mean velocity and nominal velocity dispersion of \textcolor{black}{$\Delta$\textit{v}$_{\mathrm{group}} = +210 \pm 30\ \mathrm{km\ s^{-1}}$} and \textcolor{black}{$\sigma_{\mathrm{group}} = 620 \pm 20\ \mathrm{km\ s^{-1}}$}. However, this velocity dispersion is being raised by the tail of the distribution. Consequently, the single-Gaussian velocity dispersion places an upper limit on the group mass. To place a lower limit on the quasar host group mass, we performed a second fit using \textcolor{black}{a Gaussian Mixture Model with two components.} One of these accounts for galaxies near the quasar systemic velocity while the other accounts for galaxies at \textcolor{black}{$\Delta$\textit{v} $\gtrsim +700\ \mathrm{km\ s^{-1}}$}. We found a mean velocity and nominal velocity dispersion of \textcolor{black}{$\Delta$\textit{v}$_{\mathrm{group}} = -90 \pm 20\ \mathrm{km\ s^{-1}}$} and \textcolor{black}{$\sigma_{\mathrm{group}} = 320 \pm 30\ \mathrm{km\ s^{-1}}$} for the first Gaussian, \textcolor{black}{$\Delta$\textit{v}$_{\mathrm{group}} = +1150 \pm 80\ \mathrm{km\ s^{-1}}$} and \textcolor{black}{$\sigma_{\mathrm{group}} = 330 \pm 60\ \mathrm{km\ s^{-1}}$} for the second Gaussian. \textcolor{black}{Throughout, we use jackknife resampling to estimate sampling errors on all of the Gaussian parameters, but we caution that the systematic uncertainty on the galaxy group dispersion is significantly larger due to the significant tails in the velocity distribution.} Using these two sets of Gaussian fits, the potential quasar host group dynamical mass is \textcolor{black}{$2 \times 10^{13}\ M_{\odot} \lesssim M_{\mathrm{dyn}} \lesssim 8 \times 10^{13}\ M_{\odot}$} assuming the group is virialized. \par

Galaxies in the \textcolor{black}{$\Delta$\textit{v} $= -90\ \mathrm{km\ s^{-1}}$} peak are primarily located around the quasar, \textcolor{black}{while five of the six galaxies} in the \textcolor{black}{$\Delta$\textit{v} $= +1150\ \mathrm{km\ s^{-1}}$} tail are located $d \gtrsim 50 \ \mathrm{pkpc}$ to the east of the quasar. These potential kinematic and spatial trends can be explained if, perhaps, \textcolor{black}{the smaller group of six galaxies} at \textcolor{black}{$\Delta$\textit{v} $= +1150\ \mathrm{km\ s^{-1}}$} is interacting with or infalling into \textcolor{black}{the more massive quasar host group with seventeen galaxies} at \textcolor{black}{$\Delta$\textit{v} $= -90\ \mathrm{km\ s^{-1}}$}. \textcolor{black}{We note that this interpretation is just a possibility, since this is a simple picture for a potentially more complicated scenario. If we turn the velocity difference between these two groups into a distance based on the Hubble flow, we get a line-of-sight separation of $D = 13\ \mathrm{Mpc}$\footnote{\textcolor{black}{The apparent ram pressure stripping experienced by G10 and G11 (discussed in Section \ref{sec:4.1}) suggests that cosmological effects are not responsible for the observed velocity difference.}}}. \par

To visualize the galactic environment of \pks\ in more detail, we display a 30 arcsec $\times$ 30 arcsec region of the HST WFPC2+F702W image with group members labeled in the top-left panel of Figure \ref{fig:velocity_panels}. Additionally, we display a 30 arcsec $\times$ 30 arcsec \textcolor{black}{region of the quasar light-subtracted MUSE image with nebulae labeled} in the top-middle panel of Figure \ref{fig:velocity_panels}. \textcolor{black}{This MUSE image is averaged over $6250-7250$ {\AA}, which is free of strong emission lines at $z_{\rm{QSO}} = 0.5335$.} \par

\subsection{Nebular Environment}

\textcolor{black}{The MUSE data enable the discovery of three spatially and kinematically distinct, ionized nebulae emitting strongly in [O\,II], H$\beta$, and [O\,III]. These nebulae are at projected distances of $d \approx 10-150 \ \mathrm{pkpc}$ and at LOS velocities of $\Delta$\textit{v}$ \approx -500$ to $+1500\ \mathrm{km\ s^{-1}}$ from the quasar.} Some relevant physical properties of these nebulae including size, luminosity, kinematics, and association with galaxies are shown in Table \ref{tab:summary_nebulae}. Throughout, we refer to the nebulae as labeled in the top-middle panel of Figure \ref{fig:velocity_panels} and Table \ref{tab:summary_nebulae}: Nebula 1 (N1), Nebula 2 (N2), and Nebula 3 (N3). N1 is spatially and kinematically coincident with the quasar host as well as \textcolor{black}{G1/G2/G3/G4/G5}, with $\Delta$\textit{v} $\approx +140\ \mathrm{km\ s^{-1}}$ and $\sigma \approx 140\ \mathrm{km\ s^{-1}}$. This coincidence suggest that \textcolor{black}{Host/G1/G2/G3/G4/G5} are an interacting galaxy group \textcolor{black}{responsible for} N1. N2 is spatially coincident with G10/G11/G13/G14, but with $\Delta$\textit{v} $\approx +1330\ \mathrm{km\ s^{-1}}$ and $\sigma \approx 120\ \mathrm{km\ s^{-1}}$, only G10 and G11 are kinematically coincident. This kinematic correspondence suggests that N2 arises from gas associated with G10 and G11, which are separated by $6.4 \ \mathrm{pkpc}$ in projection. G10 and G11 may be interacting with one another while also falling into the quasar host group. \textcolor{black}{Although the morphology is hard to disentangle for G10 and G11, the main reason to think they are interacting is the kinematic and spatial coincidence of these two galaxies, and association with N2.} N3 is not spatially coincident with any detected galaxies that have \textcolor{black}{secure redshifts from our redshift survey, which is approximately $90$\% complete at $m_{\mathrm{F702W}} \approx 24$.} We note, however, that there is a faint source of $m_{\mathrm{F702W}} \approx 25.4$ detected in the WFPC2 image at the eastern edge of N3, but the continuum level of the source is too low to securely measure a redshift. N3 has $\Delta$\textit{v} $\approx +550\ \mathrm{km\ s^{-1}}$ and $\sigma \approx 50\ \mathrm{km\ s^{-1}}$. We also note that G12 exhibits somewhat extended nebular emission on scales of $\approx 10 \ \mathrm{pkpc}$, but we do not consider it further in this work due to its more compact nature, and its close spatial and kinematic correspondence with G12. \par

\begin{table*}
	{\scriptsize
	\centering
	\caption{Summary of Nebulae in the Field of \pks\ at \textit{$z \approx z_{\rm{QSO}}$}.}
	\label{tab:summary_nebulae}
	\begin{threeparttable}
	\begin{tabular}{lccccccccc} 
		\hline
		ID & Area\tnote{a} & [O\,II]\tnote{b} & H$\beta$\tnote{c} & [O\,III]\tnote{d} & $\mathrm{r}_{\mathrm{[O\,II]}}$\tnote{e} & $\mathrm{r}_{\mathrm{[O\,III]}}$\tnote{f} & $\Delta$\textit{v}\tnote{g} & $\sigma$\tnote{h} & Associated Galaxies\\
		& (pkpc$^2$) & ($\mathrm{erg\ s^{-1}}$) & ($\mathrm{erg\ s^{-1}}$) & ($\mathrm{erg\ s^{-1}}$) & & & ($\mathrm{km\ s^{-1}}$) & ($\mathrm{km\ s^{-1}}$) & \\
		\hline
		N1 & \textcolor{black}{1720} & \textcolor{black}{$2.96 \times 10^{43}$} & \textcolor{black}{$8.08 \times 10^{42}$} & \textcolor{black}{$9.21 \times 10^{43}$} & $1.05 \pm 0.05$ & $0.012 \pm 0.003$ & +140 & 140 & \textcolor{black}{Host, G1, G2, G3, G4, G5}\\
		N2 & \textcolor{black}{1520} & \textcolor{black}{$1.79 \times 10^{43}$} & \textcolor{black}{$6.40 \times 10^{42}$} & \textcolor{black}{$5.79 \times 10^{43}$} & $1.36 \pm 0.05$ & $0.014 \pm 0.002$ & +1330 & 120 & G10, G11\\
		N3 & \textcolor{black}{130} & \textcolor{black}{$4.87 \times 10^{41}$} & \textcolor{black}{$1.64 \times 10^{41}$} & \textcolor{black}{$1.05 \times 10^{42}$} & $1.41 \pm 0.23$ & $0.048 \pm 0.029$ & +550 & 50 & none \\
		\hline
	\end{tabular}
	\begin{tablenotes}
	    \scriptsize
        \item \textbf{Notes.}
        \item[a] Area, with units of pkpc$^2$. 
        \item[b] Total line luminosity in [O\,II]$\lambda\lambda 3727+3729$, with units of $\mathrm{erg\ s^{-1}}$. 
        \item[c] Total line luminosity in H$\beta$, with units of $\mathrm{erg\ s^{-1}}$. 
        \item[d] Total line luminosity in [O\,III]$\lambda\lambda 4960+5008$, with units of $\mathrm{erg\ s^{-1}}$. 
        \item[e] Line ratio of [O\,II]$\rm \lambda 3729$ to [O\,II]$\rm \lambda 3727$, with errors. 
        \item[f] Line ratio of [O\,III]$\rm \lambda 4364$ to [O\,III]$\rm \lambda 5008$, with errors. 
        \item[g] LOS velocity relative to the quasar, with units of $\mathrm{km\ s^{-1}}$.
        \item[h] LOS velocity dispersion, with units of $\mathrm{km\ s^{-1}}$.
    \end{tablenotes}
	\end{threeparttable}
	}
\end{table*}

\textcolor{black}{To visualize the nebular environment of \pks\ in more detail, we display [O\,III] emission contours over the quasar light-subtracted MUSE image in the top-middle panel of Figure \ref{fig:velocity_panels}. This MUSE image is averaged over $6250-7250$ {\AA}, which is free of strong emission lines at $z_{\rm{QSO}} = 0.5335$, and allows us to understand the relationship between the morphologies of these nebulae and galaxies in the host group. Additionally, we display a LOS velocity map of the nebulae relative to the quasar in the top-right panel of Figure \ref{fig:velocity_panels}. Subsequent panels in Figure \ref{fig:velocity_panels} display narrowband images extracted from the MUSE datacube at the observed-frame wavelength of [O\,III] over the given velocities, chosen to highlight each of the nebulae and reveal detailed structure. These MUSE images allow us to understand the relationship between the kinematics of these nebulae and galaxies in the host group.} \par

\begin{figure*}
	\includegraphics[width=0.9\textwidth]{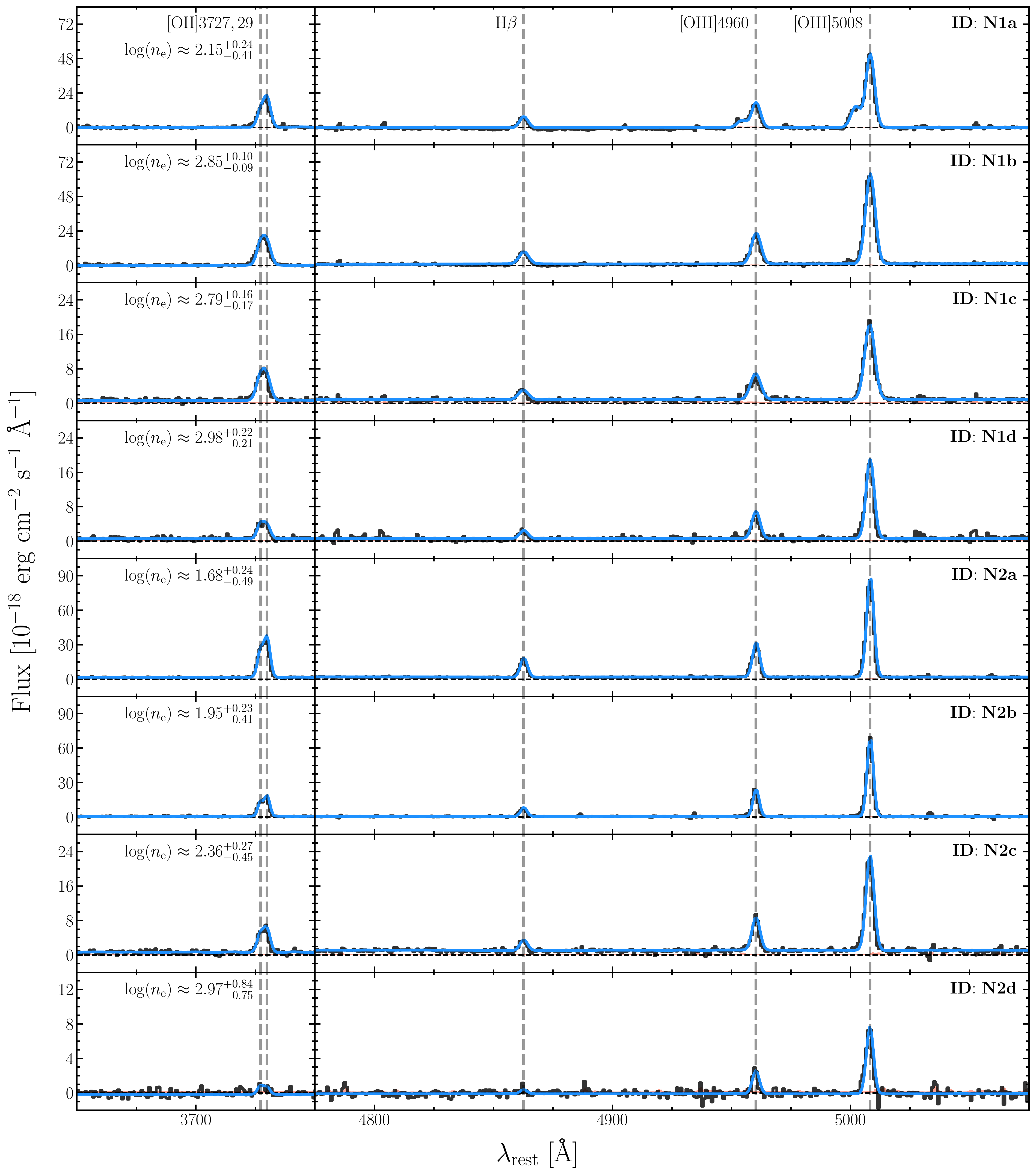}
    \caption{Examples of co-added nebular spectra and best-fitting spectral models for various regions within the two largest and most luminous nebulae (N1 and N2). The locations of these extracted regions are marked by squares and labeled by their IDs in the top-right panel of Figure \ref{fig:velocity_panels}. The MUSE spectrum is shown by a solid line in black with errors shown by a filled area in red, while the best-fit spectral model is shown by a solid line in blue. Nebula labels are in the upper-right corner of each panel. }
    \label{fig:spectrum_nebulae}
\end{figure*}

For each nebula, we measured line luminosities, LOS velocities, and LOS velocity dispersions at each spaxel using a spectral modeling technique where the continuum is fit by a linear combination of two simple stellar population models with characteristic ages of 290 Myr and 5 Gyr \citep{bruzual_stellar_2003}\textcolor{black}{\footnote{This additional continuum fit is necessary because the majority of pixels with significant nebular emission are relatively far from the nearest galaxy. Consequently, the PCA and Bagpipes stellar population synthesis models from Section \ref{sec:2.4} cannot be used to model the continuum.}}, and emission lines are fit by a single set of Gaussian line profiles that are all coupled in their radial velocity and intrinsic velocity dispersion. \textcolor{black}{To do this fit, we performed non-linear least-squares minimization using \texttt{LMFIT} \citep[][]{LMFIT} optimization techniques.} We estimated standard errors on the inferred model parameters from the MUSE error array and covariance matrix. Examples of co-added nebular spectra and best-fitting spectral models for various regions within the two largest and most luminous nebulae (N1 and N2) are shown in Figure \ref{fig:spectrum_nebulae} with two spectral windows: one covering [O\,II]$\lambda\lambda 3727,3729$, and one covering H$\beta$ and [O\,III]$\lambda\lambda 4960,5008$. The locations of these extracted regions are marked by squares and labeled by their IDs in the top-right panel of Figure \ref{fig:velocity_panels}. In addition to these strong lines, we detect several fainter lines in the brightest nebular regions, such as [Ne\,V]$\lambda 3426$, [Ne\,III]$\lambda 3869$, H$\delta$, H$\gamma$, [O\,III]$\lambda 4364$, and HeII$\,\lambda 4687$. \par 

Within the closest nebula to \pks, we noticed a second peak blueward of the [O\,III]$\lambda\lambda 4960,5008$ lines at a projected distance of $10.3 \ \mathrm{pkpc}$ from the quasar. This blue peak is visible in the top panel of Figure \ref{fig:spectrum_nebulae}, which displays the closest spectral extraction to the quasar considered here.  To account for this blue emission wing, we modified our spectral modeling technique by fitting two sets of Gaussian line components each with tied width and velocity as described previously, finding a best fit of $\Delta$\textit{v} $\approx +200\ \mathrm{km\ s^{-1}}$ from the quasar and $\sigma \approx 120\ \mathrm{km\ s^{-1}}$ for the blue wing. This sort of feature is prevalent among luminous quasars \citep{Heckman_1981, liu_outflow_2013_1, liu_outflow_2013_2} and is often associated with quasar-driven outflows.

\textcolor{black}{Using the H$\beta$ line luminosities from Table \ref{tab:summary_nebulae}, we additionally estimate the gas mass present within each of the nebulae. These measurements are made following the prescription of \citet[][]{Greene:2012}, assuming an electron density measured from the [O\,II]$\rm \lambda 3729$/[O\,II]$\rm \lambda 3727$ line ratio averaged over each nebula and Case B recombination. \textcolor{black}{This gives density estimates of $\mathrm{log}(n_{\mathrm{e}}/\mathrm{cm^{-3}}) \approx 2.8$ for N1, $\mathrm{log}(n_{\mathrm{e}}/\mathrm{cm^{-3}}) \approx 2.1$ for N2, and $\mathrm{log}(n_{\mathrm{e}}/\mathrm{cm^{-3}}) \approx 2.0$ for N3.} Additionally, we assume that the [O\,II] emitting gas density is representative of the H$\beta$ emitting gas for which we use the total line luminosity measurement\footnote{\textcolor{black}{We checked for dust extinction by measuring the H$\gamma$ to H$\beta$ line ratio within various regions extracted from each nebulae. The locations of these extracted regions are marked by squares and labeled by their IDs in the top-right panel of Figure \ref{fig:velocity_panels}. We found values consistent with Case B recombination with negligible dust extinction for each of the nebulae.}}. \textcolor{black}{We note that hydrogen and oxygen share similar ionization potentials of $13.6\ \mathrm{eV}$, so the densities measured for [O\,II] are likely representative of the regions that dominate the H$\beta$ emission (unlike the [S\,II] density diagnostic which has a lower ionization potential of $10.4\ \mathrm{eV}$; \cite{Davies:2020}).} This gives ionized mass estimates of $M_{\mathrm{g}} \approx 2.1 \times 10^{8}\ M_{\odot}$ for N1, $M_{\mathrm{g}} \approx 9.7 \times 10^{8}\ M_{\odot}$ for N2, and $M_{\mathrm{g}} \approx 2.8 \times 10^{7}\ M_{\odot}$ for N3. These ionized gas masses are lower limits on the full nebular masses because they only take into account the ionized gas that produces emission in [O\,II] and H$\beta$. Any lower density gas at higher ionization state, higher density gas at lower ionization state, or neutral gas would not be included in these estimates. These gas masses are over than an order of magnitude smaller than the cool gas masses found within the CGM of $0.21 < z < 0.55$ luminous red galaxies \citep[][]{Zahedy:2019}. The smaller masses inferred here can be explained if the nebular emission primarily arises from high density clouds with low covering fractions while the gas uncovered in CGM absorption surveys arises from lower density clouds with higher covering fractions.} \par


\section{Discussion}
\label{sec:4}

\subsection{Origin of the Nebular Gas}
\label{sec:4.1}

The two largest and most luminous nebulae (N1 and N2) are spatially and kinematically coincident with likely interacting galaxies in the field of \pks, as shown in Figure \ref{fig:velocity_panels}. In particular, N1 spatially and kinematically surrounds \textcolor{black}{Host/G1/G2/G3/G4/G5}  while N2 spatially and kinematically surrounds the interacting galaxy pair G10/G11. Morphologically, N2 has a head-tail structure extending $d \approx 85 \ \mathrm{pkpc}$ to the northeast, and a velocity gradient with $\Delta$\textit{v} $\approx +1500\ \mathrm{km\ s^{-1}}$ at the head of N2 and $\Delta$\textit{v} $\approx +800\ \mathrm{km\ s^{-1}}$ at the tail. This spatial and kinematic extension is suggestive of ram pressure stripping that would result if G10 and G11 are moving through \textcolor{black}{the hot halo associated with the galaxy over-density. Ram pressure stripping can be effective in groups with masses of $M_{\rm dyn}\approx 10^{13}-10^{14}$ as demonstrated by the existence of ``jellyfish'' galaxies and ram pressure stripped nebulae in group environments \citep[e.g.][]{Poggianti:2016, chen_giant_2019}} The \textcolor{black}{dynamical and spatial association} with \textcolor{black}{likely} interacting galaxies, the narrow internal velocity dispersions exhibited by these nebulae ($\sigma \approx 110 - 140\ \mathrm{km\,s^{-1}}$), and the high densities of these nebulae ($n_{\mathrm{e}} \approx 60 - 800\ \mathrm{cm^{-3}}$) \textcolor{black}{all suggest that these are stripped ISM from ongoing interactions} \citep{poggianti_gasp._2017, bellhouse_gasp._2017}. \par

Unlike the other nebulae, N3 is a narrow (both in velocity and morphology) gas cloud located $d \approx 70-95 \ \mathrm{pkpc}$ from the quasar with a width of $\lesssim 10 \ \mathrm{pkpc}$, a LOS velocity of $\Delta$\textit{v} $\approx +550\ \mathrm{km\ s^{-1}}$, and a LOS velocity dispersion of $\sigma \approx 50\ \mathrm{km\ s^{-1}}$ from the quasar, as shown in the bottom-left panel of Figure \ref{fig:velocity_panels}. N3 is the faintest of the three nebulae with the lowest surface brightness, the lowest internal velocity dispersion, and \textcolor{black}{slightly lower ionization}. While the eastern edge of N3 is coincident with a faint source of $m_{\mathrm{F702W}} \approx 25.4$, it is not associated with any detected galaxies that have secure redshifts. The morphology, low surface brightness, calm kinematics, low ionization state, and lack of any associated galaxies \textcolor{black}{are all consistent with this being a cool gas structure in the IGrM} \citep{borthakur_2010, steidel_structure_2010, johnson_galaxy_2018, chen_giant_2019}, or gas associated with a dwarf galaxy fainter than $M_{g} \approx -18.5$ (at $z_{\rm{QSO}} = 0.5335$). \par

\subsection{Origin of the Nebular Photoionization}

From Table \ref{tab:summary_nebulae} and Figure \ref{fig:spectrum_nebulae}, we see that the nebulae exhibit high ionization states with mean log([O\,III]$\rm \lambda 5008$/[O\,II]$\rm \lambda \lambda 3727,3729$) $\approx 0.3 - 0.6$ and mean log([O\,III]$\rm \lambda 5008$/$\rm H \beta$) $\approx 0.9 - 1.3$. \textcolor{black}{These line ratios} can be produced with photoionization by a quasar \citep{groves_dusty_2004-2} or fast shocks \citep{allen_mappings_2008}, but cannot be produced with photoionization by stars, even from a very young stellar population \citep{morisset_virtual_2014}. \par

To better quantify the ionization properties of the gas, we compare various line ratios (${\rm [O\,III]\lambda 5008 / \rm [O\,II]\lambda \lambda 3727,3729}$, ${\rm [O\,III]\lambda 5008 / \rm H \beta}$, ${\rm [O\,III]\lambda 4364 / \rm [O\,III]\lambda 5008}$, ${\rm [Ne\,V]\lambda 3427 / \rm [Ne\,III]\lambda 3869}$) for the brightest nebular regions with dusty radiation pressure-dominated quasar models, calculated with \texttt{MAPPINGS III} \citep{groves_dusty_2004-2, groves_dusty_2004-1} and fully radiative shock models, calculated with \texttt{MAPPINGS V} \citep{sutherland_effects_2017}. We also compare these line ratios for the brightest nebular regions with dusty radiation pressure-dominated photoionization models from \texttt{Cloudy} \citep{ferland_Cloudy_2017}. Figure \ref{fig:_line_diagnostics} shows these comparisons. We only consider nebular regions with signal-to-noise ratio $\mathrm{S/N} > 3$ in the strong lines for this analysis, unless stated otherwise. \par

For the \texttt{MAPPINGS III} models we use the IDL Tool for Emission-line Ratio Analysis (\texttt{ITERA}), which assumes \textcolor{black}{the spectrum of the ionizing source to be a simple power law}, abundances similar to those deduced by \cite{allen_chemodynamical_1998}, and an electron density motivated by observations \citep{groves_itera:_2010}. For the \texttt{MAPPINGS V} models we use the Mexican Million Models dataBase-shocks \citep[\texttt{3MdBs}][]{alarie_extensive_2019}, which assumes a shock plus precursor model and abundances from \citet{allen_mappings_2008}. For the \texttt{Cloudy} models we assume different incident quasar radiation field shapes, chemical compositions, and density laws. The \texttt{MAPPINGS III} model grids span various power law slopes ($\alpha$) and ionization parameters ($U$) while the \texttt{MAPPINGS V} model grids span various shock velocities ($v_{\mathrm{shock}}$) and magnetic field strengths ($B$). The ionization parameter $U$ is defined here as the dimensionless ratio of hydrogen-ionizing photons to total-hydrogen densities. \par

We use the [O\,II]$\rm \lambda 3729$/[O\,II]$\rm \lambda 3727$ line ratio averaged over each nebula to measure the electron density ($n_{\mathrm{e}}$) and the [O\,III]$\rm \lambda 4364$/[O\,III]$\rm \lambda 5008$ line ratio averaged over each nebula to measure the electron temperature ($T_{\mathrm{e}}$) following the prescription of \citet{Osterbrock:2006} as implemented in \texttt{PyNeb} \citep{luridiana_pyneb:_2015}. These line ratios are given in Table \ref{tab:summary_nebulae} and imply values of \textcolor{black}{$\mathrm{log}(n_{\mathrm{e}}/\mathrm{cm^{-3}}) \approx 2.8 \pm 0.2$} and \textcolor{black}{$\mathrm{log}(T_{\mathrm{e}}/\mathrm{K}) \approx 4.1 \pm 0.1$} for the nebular regions associated with N1 and \textcolor{black}{$\mathrm{log}(n_{\mathrm{e}}/\mathrm{cm^{-3}}) \approx 2.1 \pm 0.3$} and \textcolor{black}{$\mathrm{log}(T_{\mathrm{e}}/\mathrm{K}) \approx 4.1 \pm 0.1$} for the nebular regions associated with N2, where the uncertainties reflect intrinsic scatter in each nebula. The S/N in both the [O\,II] and [O\,III] lines for N3 is too low to accurately measure the electron density or temperature. \par

The three spatially and kinematically distinct, ionized nebulae exhibit very different galaxy associations, LOS velocities, projected physical distances, and electron density. These different projected physical distances and electron densities necessitate separate photoionization modeling. While N1 and N2 are strongly emitting, allowing for significant detection in many of the emission lines used for this analysis and accurate measurements of electron density and temperature, N3 is weakly emitting with no significant detection in many of the emission lines used for this analysis and no accurate measurements of electron density and temperature. Thus, we will only consider N1 and N2 for subsequent photoionization analysis. In Figure \ref{fig:_line_diagnostics}, nebular regions associated with N1 are shown by the red points, while nebular regions associated with N2 are shown by the blue points. \par 

\subsubsection{Nebula 1: Close to the Quasar}

For the nebular regions associated with N1, we find that the \texttt{MAPPINGS III} quasar model with solar metallicity, $-3.3\ \leq \ \mathrm{log}(U)\ \leq \ -2.3$, and $-2.0\ \leq \ \alpha\ \leq \ -1.2$ is consistent with the observed strong line ratios, as shown by the red points in the left column of Figure \ref{fig:_line_diagnostics}.  The range in ionization parameter shown is the full physically realizable range based on \textcolor{black}{the observed monochromatic UV luminosity of the quasar observed at rest-frame 0.7 Rydberg in the archival COS/FUV spectrum (PI: C. Churchill, PID: 12252)}, the measured electron densities of these nebulae ($\mathrm{log}(n_{\mathrm{e}}/\mathrm{cm^{-3}}) \approx 2.8 \pm 0.2$), and the various projected distances of these nebulae ($d \approx 10-75 \ \mathrm{pkpc}$). \textcolor{black}{We adopted the UV monochromatic luminosity measured with COS instead of the bolometric luminosity given in Section 2.2 to prevent bolometric correction uncertainty from impacting the photoionization analysis.} \par

While we are able to reproduce the observed strong line ratios (${\rm [O\,III]\lambda 5008 / \rm [O\,II]\lambda \lambda 3727,3729}$ and ${\rm [O\,III]\lambda 5008 / \rm H \beta}$) and the temperature sensitive weak line ratio (${\rm [O\,III]\lambda 4364 / \rm [O\,III]\lambda 5008}$) with this model, we are unable to reproduce the ionization sensitive weak line ratio (${\rm [Ne\,V]\lambda 3427 / \rm [Ne\,III]\lambda 3869}$). \textcolor{black}{In order to explain the ionization sensitive line ratio (${\rm [Ne\,V]\lambda 3427 / \rm [Ne\,III]\lambda 3869}$), the quasar would need to be at least $10$ times brighter, with $L_{\rm{bol}} \gtrsim 10^{48}\ \rm{erg\ s^{-1}}$ using the gas density inferred from the [O\,II] doublet. Given the rarity of such hyperluminous quasars at $z < 1$, this suggests that the [Ne\,V] emission may arise from a lower density gas phase than the [O\,II] and [O\,III] emission \citep[][]{Davies:2020}} \par

On the other hand, the \texttt{MAPPINGS V} shock model with  solar metallicity, $200\ \mathrm{km\,s^{-1}}\ \leq \ v_{\mathrm{shock}}\ \leq \ 600\ \mathrm{km\,s^{-1}}$, and $1.0\ \mathrm{\mu G}\ \leq \ B\ \leq \ 100.0\ \mathrm{\mu G}$ are inconsistent with the measured strong and weak line ratios, as shown by the red points in the right column of Figure \ref{fig:_line_diagnostics}. \par

For the nebular regions considered here, \textcolor{black}{there is generally an anti-correlation between the line ratios [O\,III]$\rm \lambda 5008$/[O\,II]$\rm \lambda \lambda 3727,3729$ and the projected distance from the quasar ([O\,III]$\rm \lambda 5008$/[O\,II]$\rm \lambda \lambda 3727,3729 \approx 0.5$ at $d \approx 10\ \mathrm{pkpc}$, [O\,III]$\rm \lambda 5008$/[O\,II]$\rm \lambda \lambda 3727,3729 \approx 0.3$ at $d \approx 20\ \mathrm{pkpc}$)}, consistent with gas in the quasar host group that is photoionized by the quasar. Additionally, the internal velocity dispersions are low with $\sigma \approx 140\ \mathrm{km\,s^{-1}}$, which disfavors the shock scenario since it requires relatively fast shocks of $v_{\mathrm{shock}} \approx 200 - 600\ \mathrm{km\,s^{-1}}$ to reproduce the observed line ratios. Based on this and the red points of Figure \ref{fig:_line_diagnostics}, it seems that the nebular regions associated with N1 are predominantly being photoionized by the quasar. \par

To try and reproduce the other observed weak line ratio, we compared the ionization properties of the gas with dusty radiation pressure-dominated photoionization models from \texttt{Cloudy} \citep{ferland_Cloudy_2017} assuming different incident radiation field shapes, chemical compositions, and densities. With the additional model parameters explored by this technique, we were able to reproduce some of the observed line ratios (${\rm [O\,III]\lambda 5008 / \rm [O\,II]\lambda \lambda 3727,3729}$, ${\rm [O\,III]\lambda 5008 / \rm H \beta}$, and ${\rm [O\,III]\lambda 4364 / \rm [O\,III]\lambda 5008}$), but not all (${\rm [Ne\,V]\lambda 3427 / \rm [Ne\,III]\lambda 3869}$). In future work, we will explore additional possibilities including log-normal density distributions \citep{cantalupo_2019} which may explain the more highly ionized emission. Nevertheless, based on the proximity to the quasar, strong-line ratios, anti-correlation between [O\,III]$\rm \lambda 5008$/[O\,II]$\rm \lambda \lambda 3727,3729$ and projected distance, and kinematics, it seems that the nebular regions associated with N1 are predominantly being photoionized by the quasar. \par

\subsubsection{Nebula 2: Far from the Quasar}

\begin{figure*}
	\includegraphics[width=0.67\textwidth]{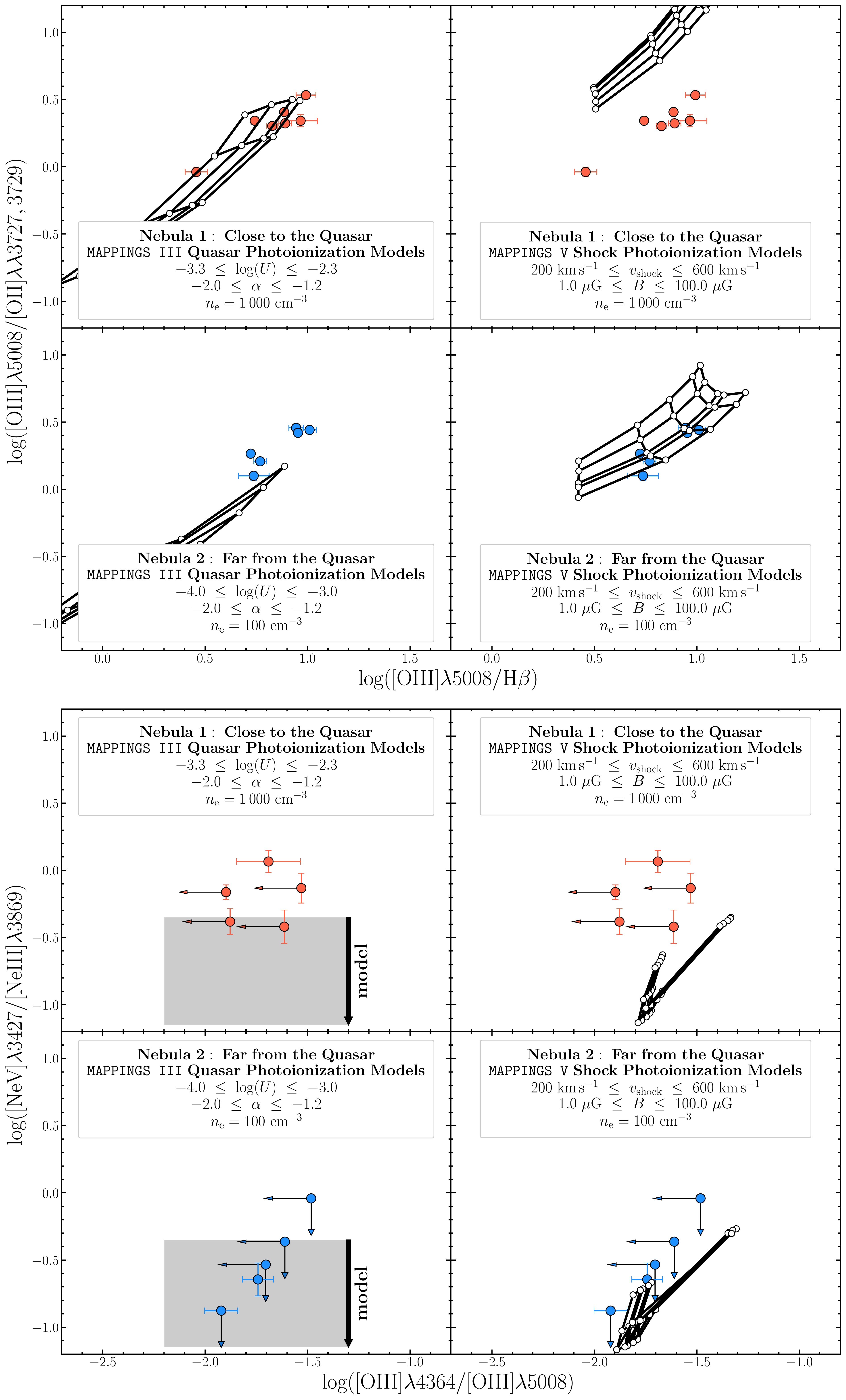}
    \caption{The emission line diagnostic diagrams log([O\,III]$\rm \lambda 5008$/[O\,II]$\rm \lambda \lambda 3727,3729$) vs. log([O\,III]$\rm \lambda 5008$/$\rm H \beta$) and ${\rm log (\rm [Ne\,V]\lambda 3427 / \rm [Ne\,III]\lambda 3869)}$ vs. ${\rm log (\rm [O\,III]\lambda 4364 / \rm [O\,III]\lambda 5008)}$ for the various extended line-emitting nebulae.  Nebular regions associated with N1 are shown by the red points and labeled in the legend, while nebular regions associated with N2 are shown by the blue points and labeled in the legend. For nebulae with $\mathrm{S/N} > 3$ in each of the given lines, we plot points with error bars. Otherwise, we plot $3\sigma$ upper limits. We compare these line ratios with dusty radiation pressure-dominated quasar models calculated with \texttt{MAPPINGS III} \citep{groves_dusty_2004-2, groves_dusty_2004-1} in the left column and fully radiative shock models calculated with \texttt{MAPPINGS V} \citep{sutherland_effects_2017} in the right column.  For the \texttt{MAPPINGS III} models we assume a simple power law to represent the spectrum of the ionizing source, abundances similar to that deduced by Allen et al. (1998), $n_{\mathrm{e}} = 1\,000\ \mathrm{cm^{-3}}$ for the nebular regions associated with N1, and $n_{\mathrm{e}} = 100\ \mathrm{cm^{-3}}$ for the nebular regions associated with N2.  For the \texttt{MAPPINGS V} models we assume a shock plus precursor model, abundances similar to that deduced by Allen et al. (2008), $n_{\mathrm{e}} = 1\,000\ \mathrm{cm^{-3}}$ for the nebular regions associated with N1, and $n_{\mathrm{e}} = 100\ \mathrm{cm^{-3}}$ for the nebular regions associated with N2.  The \texttt{MAPPINGS III} model grids span various power law slopes ($\alpha$) and ionization parameters ($U$) while the \texttt{MAPPINGS V} model grids span various shock velocities ($v_{\mathrm{shock}}$) and magnetic field strengths ($B$), as shown in the legend. }
    \label{fig:_line_diagnostics}
\end{figure*}

For the nebular regions associated with N2, we find that the \texttt{MAPPINGS III} quasar model with solar metallicity, $-4.0\ \leq \ \mathrm{log}(U)\ \leq \ -3.0$, and $-2.0\ \leq \ \alpha\ \leq \ -1.2$ is inconsistent with the observed line ratios, as shown by the blue points in the left column of Figure \ref{fig:_line_diagnostics}. The range in ionization parameter quoted here is the full physically realizable range based on the observed luminosity of the quasar, the measured electron densities of these nebulae ($\mathrm{log}(n_{\mathrm{e}}/\mathrm{cm^{-3}}) \approx 2.1 \pm 0.3$), and the various projected distances of these nebulae ($d \approx 75-150 \ \mathrm{pkpc}$). We are unable to reproduce either the observed strong line ratios or the observed weak line ratios with this model, given the full physically realizable range in ionization parameter. \par

Furthermore, we compared the ionization properties of the gas with dusty radiation pressure-dominated photoionization models from \texttt{Cloudy} \citep{ferland_Cloudy_2017} assuming many different incident radiation field shapes, chemical compositions, and densities. We find that these conclusions about the origin of the nebular photoionization still hold, and we do not gain any more information about the gas, even with these additional free parameters. In order to explain the observed line ratios, the quasar would need to have been $\approx 10$ times brighter in the past, with $L_{\rm{bol}} \approx 10^{48}\ \rm{erg\ s^{-1}}$ roughly $0.3 \ \mathrm{Myr}$ ago. \par 

We find that the \texttt{MAPPINGS V} shock model with solar metallicity, $200\ \mathrm{km\,s^{-1}}\ \leq \ v_{\mathrm{shock}}\ \leq \ 600\ \mathrm{km\,s^{-1}}$, and $1.0\ \mathrm{\mu G}\ \leq \ B\ \leq \ 100.0\ \mathrm{\mu G}$ to be consistent with the measured strong and weak line diagnostics, as shown by the blue points in the right column of Figure \ref{fig:_line_diagnostics}. \par 

\textcolor{black}{Another potential source of the photoionization for the nebular regions considered here is an obscured AGN hosted by G10/G11. However, the line ratios [O\,III]$\rm \lambda 5008$/[O\,II]$\rm \lambda \lambda 3727,3729$ for these nebular regions are generally correlated with the projected distance from G10/G11 ([O\,III]$\rm \lambda 5008$/[O\,II]$\rm \lambda \lambda 3727,3729 \approx 0.3$ at $d \approx 0\ \mathrm{pkpc}$, [O\,III]$\rm \lambda 5008$/[O\,II]$\rm \lambda \lambda 3727,3729 \approx 0.5$ at $d \approx 20\ \mathrm{pkpc}$), while one would expect an anti-correlation if the ionizing photons came from an obscured AGN in G10 or G11. Additionally, spectra extracted from a narrow region around G10/G11 (in order to prevent nebular contamination) do not have significantly detected ${\rm [Ne\,V]\lambda 3427}$ emission, which is commonly used as a tracer for gas photoionized by an AGN \citep{Mignoli:2013, Feltre:2016}. Typical obscured AGN at $0.3 < z < 0.8$ exhibit [Ne\,V]$\rm \lambda 3427$/[O\,II]$\rm \lambda \lambda 3727,3729 \approx 0.2$ \citep{Zakamska:2003}, while G10/G11 exhibit a 3$\sigma$ upper limit of [Ne\,V]$\rm \lambda 3427$/[O\,II]$\rm \lambda \lambda 3727,3729 \lesssim 0.08$. For these reasons, it seems unlikely that that G10/G11 hosts an obscured AGN responsible for the photoionzation of N2. For similar reasons, we do not expect there to be an obscured AGN hosted by G13/G14.} \par

While the nebular regions considered here have low internal velocity dispersions with $\sigma \approx 40 - 140\ \mathrm{km\,s^{-1}}$, they are likely moving through the IGrM of some galaxy over-density around \pks\ with $+800\ \mathrm{km\ s^{-1}}$ $\lesssim$ $\Delta$\textit{v} $\lesssim$ +1500 $\mathrm{km\ s^{-1}}$,  comparable to the required shock velocities from the models.  Based on this and Figure \ref{fig:_line_diagnostics}, it seems that the nebular regions associated with N2 are predominantly being photoionized by fast shocks, which could arise from the interaction between nebulae and the hot IGrM of some galaxy over-density.\par

\section{Summary and Conclusions}
\label{sec:5}

We reported wide-field optical integral field spectroscopy of \pks, \textcolor{black}{which is among the the most luminous 0.1\% quasars in the $z < 1$ Universe \citep{Shen:2011},} with MUSE at the VLT. The wide field of view, high spatial sampling, and broad wavelength coverage of MUSE allowed us to thoroughly study the gaseous and group environment hosting \pks\ using a spatially resolved analysis of the given morphologies, kinematics, and nebular photoionization properties. Our findings can be summarized as follows.

\begin{enumerate}
    \item We found that \pks\ sits in an overdense region of galaxies comprised of two potential kinematic and spatial groupings. This can be explained if perhaps a smaller group of \textcolor{black}{six} galaxies at \textcolor{black}{$\Delta$\textit{v} $\approx +1150\ \mathrm{km\ s^{-1}}$} and \textcolor{black}{$\sigma \approx 330\ \mathrm{km\ s^{-1}}$} is interacting with or infalling into the more massive quasar host group with \textcolor{black}{seventeen} galaxies at \textcolor{black}{$\Delta$\textit{v} $\approx -90\ \mathrm{km\ s^{-1}}$} and with \textcolor{black}{$\sigma \approx 320\ \mathrm{km\ s^{-1}}$}. \textcolor{black}{The potential dynamical mass of the host group is $2 \times 10^{13}\ M_{\odot} \lesssim M_{\mathrm{dyn}} \lesssim 8 \times 10^{13}\ M_{\odot}$ assuming the group is virialized.}
    \item From these galaxies, we found an interacting galaxy group (\textcolor{black}{Host/G1/G2/G3/G4/G5}, see Figure \ref{fig:HST_image}) with a projected separation of $37.4 \ \mathrm{pkpc}$ and an interacting galaxy pair (G10/G11, see Figure \ref{fig:HST_image}) with a projected separation of $6.4 \ \mathrm{pkpc}$. These interacting groups were determined based on galaxy morphologies, projected separations, relative velocities, and nebular coincidence.
    \item We identified three spatially and kinematically distinct, ionized nebulae emitting strongly in [O\,II], H$\beta$, and [O\,III] on scales of $10-150 \ \mathrm{pkpc}$ and at LOS velocities of $\Delta$\textit{v} $\approx -500$ to $+1500\ \mathrm{km\ s^{-1}}$ from the quasar. In particular, the first is spatially coincident with the interacting galaxy group (N1, see Figure \ref{fig:velocity_panels}), the second is spatially coincident with the interacting galaxy pair (N2, see Figure \ref{fig:velocity_panels}), and the third is not spatially coincident with any galaxies that have secure redshifts (N3, see Figure \ref{fig:velocity_panels}). 
    \item We found that the two largest and most luminous nebulae (N1 and N2) \textcolor{black}{are most likely stripped ISM from ongoing interactions} based on their \textcolor{black}{dynamical and spatial association with likely interacting galaxies}, their narrow internal velocity dispersions, and their high densities. On the other hand, we found that N3 is most likely a cool gas structure in the intragroup medium based on its morphology, low surface brightness, calm kinematics, low ionization state, and lack of any associated galaxies that have secure redshifts. 
    \item To better understand the two largest and most luminous nebulae (N1 and N2), we compared optical line ratios for some of the brightest nebular regions with dusty radiation pressure-dominated quasar photoionization models and radiative shock models to better quantify the ionization properties of the gas. We found that the nebular regions associated with N1 are most likely being predominantly photoionized by the quasar. We found that the nebular regions associated with N2 are most likely being predominantly photoionized by fast shocks, which could arise from the interaction between nebulae and the hot intragroup medium of the quasar host group.  
\end{enumerate}

Using wide-field optical integral field spectroscopy, \textcolor{black}{we were able to simultaneously survey the galactic and gaseous environment around \pks\ which allowed for (1) sensitive observations of extended line-emitting nebulae and (2) joint studies of their morphologies and kinematics in the context of those of nearby galaxies.} The presence of these spatially extended ionized nebulae in the host group of \textcolor{black}{one of the most luminous 0.1\% quasars in the $z < 1$ Universe \citep{Shen:2011}} provide evidence supporting interaction related triggering of quasars. These interactions also drive large quantities of interstellar medium to the circumgalactic medium, producing cool gas structures on halo scales. This demonstrates the power of integral field spectroscopy when combined with photoionization from active galactic nuclei and shocks for revealing large-scale gas supplies in nebular emission lines. While focusing on a single system such as \pks\ is beneficial, we need further analysis on other systems and robust statistics to fully understand the importance of galaxy environment and interactions in fueling galaxy and black hole growth.

\newpage
\section*{Acknowledgements}
We thank the the referee for providing constructive and insightful suggestions which helped strengthen the paper. We are grateful to C. P\'eroux both for helpful comments on the paper draft as well as leadership of the MUSE program that obtained the archival data that enabled our study. We thank M. Gaspari, R. Cen, and M. Strauss for fruitful discussions of this project. SDJ is grateful for partial support from a NASA Hubble Fellowship (HST-HF2-51375.001-A) and a Carnegie-Princeton Fellowship. JEG acknowledges support from a National Science Foundation grant (AST-1815417). This paper is based on observations from the European Organization for Astronomical Research in the Southern Hemisphere under ESO (PI: C. P\'eroux, PID: 0100.A-0753) and the NASA/ESA Hubble Space Telescope (PI: K. Lanzetta, PID: 7475; PI: C. Churchill, PID: 12252). Additionally, this paper made use of the NASA/IPAC Extragalactic Database, the NASA Astrophysics Data System, \texttt{Astropy} \citep[][]{Astropy-Collaboration:2018}, and \texttt{Aplpy} \citep[][]{Robitaille:2012}.

\section*{Data availability}
The data used in this paper are all publicly available from the ESO (PI: C. P\'eroux, PID: 0100.A-0753) and HST data archives (PI: K. Lanzetta, PID: 7475; PI: C. Churchill, PID: 12252).



\bibliographystyle{mnras}
\bibliography{references}
\newpage



\appendix
\section{Quasar Light Subtraction}
\label{sec:6}

Luminous quasars significantly outshine their host galaxies, and at intermediate redshifts ($z \gtrsim 0.5$) quasar light can contaminate neighboring galaxies in ground-based data. Mitigating this effect in order to gain useful information about the quasar host galaxy, as well as neighboring galaxies, requires subtraction of the quasar light contamination. Traditional approaches for quasar light subtraction often use an empirical determination of the point-spread function (PSF) based on nearby, bright and unsaturated stars \citep{kamann_resolve_2013}. However, not all observations contain a nearby, bright and unsaturated star within the 1.0 arcmin $\times$ 1.0 arcmin field of view (FOV) of MUSE. Alternative approaches for quasar light subtraction at higher redshifts ($z \gtrsim 2$) adopt the quasar as the PSF, with the assumption that there is no contribution from the quasar host galaxy or neighboring galaxies \citep{cantalupo_2019}. However, at lower redshifts ($z \lesssim 1$) where quasar hosts and neighbors contribute non-negligible extended flux, this approach will lead to over-subtraction. \par

\begin{figure}
    \centering
	\includegraphics[width=1.0\linewidth]{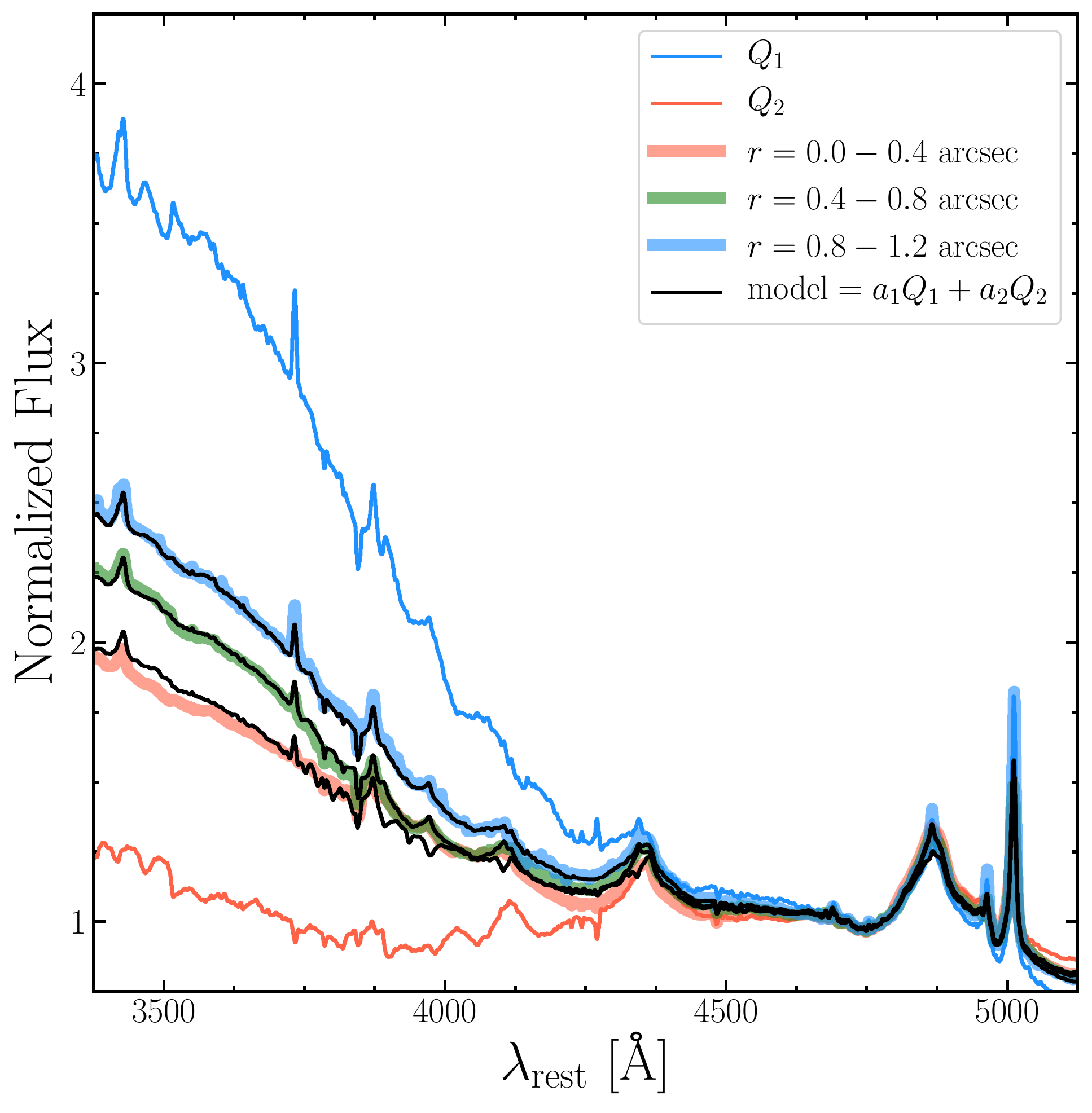}
    \caption{Examples of co-added quasar spectra extracted in annuli and best-fitting spectral models from a linear combination of NMF components. The annuli are centered on the quasar centroid with inner and outer radii specified in the figure legend. Due to the wavelength dependent seeing, the spectra extracted from smaller annuli are artificially shallow while the spectra extracted from larger annuli are artificially steep. To model this effect, we performed NMF on spaxels dominated by the quasar light to define two non-orthogonal, non-negative spectral components ($Q_{1}$ and $Q_2$) which are able to approximate the quasar contribution as a linear combination. To demonstrate the effectiveness of this technique, we fit each annular extracted spectrum with a linear combination of the two non-orthogonal, non-negative spectral components ($a_1 Q_1 + a_2 Q_2$) and overplot the best-fits in black. }
    \label{fig:appendix1}
\end{figure}

\textcolor{black}{To overcome these challenges, \citet{johnson_galaxy_2018} developed a quasar light subtraction technique that is described here in more detail. This technique is free of assumptions about the PSF shape. Instead, it uses spectral information from the integral field spectrograph (IFS) and the fact that galaxy and quasar spectral energy distributions are significantly different \citep[for similar approaches, see][]{husemann_quasar_properties_2013, rupke_quasar-mode_2017}.} Due to the wavelength dependence of atmospheric seeing, blue light from the quasar is dispersed further away from the quasar centroid than red light. This effect causes the contaminating spectrum from the quasar to be artificially flat (as a function of wavelength) close to the quasar, and artificially steep far from it. Figure \ref{fig:appendix1} demonstrates this effect by comparing co-added quasar spectra extracted in annuli around the quasar, with the spectra becoming steeper for larger annuli. To account for this wavelength dependence, we used non-negative matrix factorization (NMF) \citep{blanton_k_2007, ren_non-negative_2018} to derive two non-orthogonal, non-negative spectral components that can describe the quasar contamination to any spaxel as a linear combination. These two non-orthogonal, non-negative spectral components are defined by performing NMF on quasar-dominated spaxels from a 1.0 arcsec $\times$ 1.0 arcsec aperture centered around the quasar. The first of these components ($Q_1$, as shown by the blue line in Figure \ref{fig:appendix1}) has a steep spectral slope and approximates quasar contamination to spaxels far from the quasar. The second of these components ($Q_2$, as shown by the red line in Figure \ref{fig:appendix1}) has a shallow spectral slope and approximates quasar contamination to spaxels close to the quasar. Linear combinations of these two spectral components accurately model quasar light contributions at $r \gtrsim 0.4$ arcsec from the quasar centroid (as shown by the black lines in Figure \ref{fig:appendix1}). \par

\begin{figure*}
    \centering
	\includegraphics[width=1.0\linewidth]{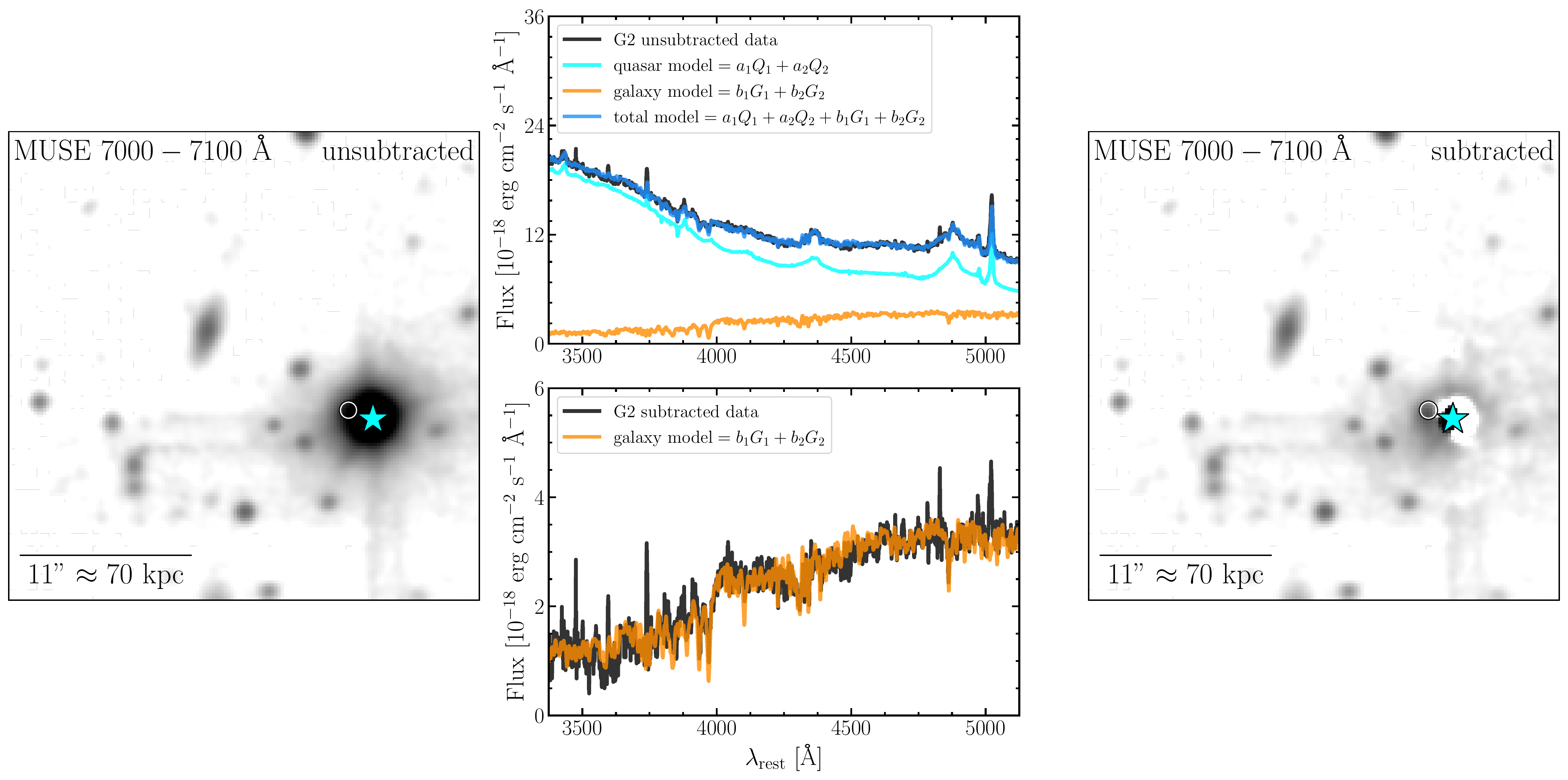}
    \caption{Example of the quasar light subtraction technique. A slice of the unsubtracted MUSE datacube is shown in the left panel while a spectrum from a nearby galaxy at 11.0 arcsec from the quasar (G2) is shown in the top middle panel. We fit the spectrum with a linear combination of galaxy eigenspectra from BOSS, before subtraction. A slice of the subtracted MUSE datacube is shown in the right panel while a spectrum from the closest galaxy to the quasar is shown in the bottom middle panel. We fit the spectrum with a linear combination of galaxy eigenspectra from BOSS, before subtraction. Due to the wavelength dependent seeing, the blue end of the spectrum for this galaxy contains significant contributions from quasar light. }
    \label{fig:appendix2}
\end{figure*}

To subtract quasar light from the contaminated regions, we modeled each spaxel at $r \lesssim$ 6 arcsec from the quasar as a linear combination of $Q_1$ and $Q_2$, as well as \textcolor{black}{the first} two galaxy eigenspectra from the Baryon Oscillation Spectroscopic Survey \citep[BOSS;][]{bolton_spectral_2012}, shifted to $z_{\rm{QSO}} = 0.5335$ with the strongest emission lines masked. In particular, the spectrum at each location is modeled by Equation \ref{eq:A1}
\begin{equation}
    \label{eq:A1}
    m(\lambda) = a_1 Q_1 (\lambda) + a_2 Q_2 (\lambda) + b_1 G_1 (\lambda) + b_2 G_2 (\lambda)
\end{equation}
where $G_1,\,G_2$ are the BOSS galaxy eigenspectra and $a_1,\,a_2,\,b_1,\,b_2$ are free parameters. We then subtracted the quasar component of the best fit model ($a_1 Q_1 + a_2 Q_2$) from each spaxel. \textcolor{black}{This effectively removes the quasar light contribution at $\gtrsim$ 1 arcsec from the quasar.} At $r \lesssim$ 1 arcsec from the quasar centroid, the residuals are significant as seen in the right panel of Figure \ref{fig:appendix2}, so we mask this area in all measurements presented here. This modeling and subtraction is demonstrated in the middle two panels of Figure \ref{fig:appendix2} for a nearby galaxy at 11.0 arcsec from the quasar (G2). \par

\label{lastpage}
\end{document}